\documentclass[final,3p,times] {elsarticle}

\usepackage{amssymb}

\usepackage[]{amsmath}
\usepackage{graphics}
\usepackage{amssymb}
\usepackage{amsthm}
\usepackage{setspace}
\usepackage{epsfig}
\usepackage{subfigure}
\usepackage{booktabs}
\usepackage{epstopdf}
\usepackage{float}
\usepackage{mathrsfs}

\biboptions{numbers,sort&compress}
\usepackage[pdftex,colorlinks,linkcolor=red,anchorcolor=blue,citecolor=green]{hyperref}

\usepackage{amsmath}
\usepackage{times}
\usepackage{anysize}
\marginsize{2.5cm}{2.5cm}{1cm}{2cm}

\selectfont

\begin{document}

\begin{frontmatter}

\newtheorem{theorem}{Theorem}
\newtheorem{lemma}[theorem]{Lemma}

\title{A deep learning method for solving high-order nonlinear soliton equation}

\author[label1]{ Shikun Cui  \corref{cor2}}
\author[label1,label2]{ Zhen Wang \corref{cor1}}\ead{wangzhen@dlut.edu.cn}
\author[label1]{ Jiaqi Han  \corref{cor2}}
\author[label1]{ Xinyu Cui  \corref{cor2}}
\cortext[cor1]{School of Mathematical Sciences, Dalian University of Technology, Dalian, 116024, China}
\address[label1]{School of Mathematical Sciences, Dalian University of Technology, Dalian, 116024, China}
\address[label2]{Key Laboratory for Computational Mathematics and Data Intelligence of Liaoning Province, Dalian, 116024, China}

\begin{abstract}
We propose effective scheme of deep learning method for high-order nonlinear soliton equation and compare the activation function for high-order soliton equation. The neural network approximates the solution of the equation under the conditions of differential operator, initial condition and boundary condition. We apply this method to high-order nonlinear soliton equation, and verify its efficiency by solving the fourth-order Boussinesq equation and the fifth-order Korteweg de Vries equation. The results show that deep learning method can solve the high-order nonlinear soliton equation and reveal the interaction between solitons.
\end{abstract}

\begin{keyword}
deep learning method, physics-informed neural networks, high-order nonlinear soliton equation, interaction between solitons, numerical driven solution.
\end{keyword}
\end{frontmatter}
\section{Introduction}

Nonlinear soliton equation is an important part of the field of Mathematical Physics, which is used to describe the state or process changing with time in physics, mechanics or other natural sciences\cite{MayersJ-2008}\cite{NenadManojlovic-1995}. As the carrier of soliton theory, the development of nonlinear equation has always been the focus of mathematical physics researchers. In recent years, a deep learning numerical method has been developed to solve many problems related to nonlinear evolution equation. Deep learning method approximate potential solutions by  using deep neural network , which is usually more effective than ordinary numerical methods\cite{MRaissi-2017}\cite{Lagaris-1998}\cite{RaissiM-2018}. Raissi M et al\cite{MRaissi-2017} proposed physics-informed neural networks to solve partial differential equation, such neural networks are constrained to respect any symmetries, invariances, or conservation principles. Han et al\cite{JHan-2018} reconstruct partial differential equations from backward stochastic differential equations, and use neural networks to approximate the gradient of unknown solutions to solve general high-dimensional parabolic partial differential equations. Justin et al\cite{JSirignano-2018} used DGM( depth Galerkin method) to study the numerical driven solution of high dimensional partial differential equation. Li and Chen\cite{LJCY0}\cite{LJCY1} used deep learning method to solve many second-order nonlinear evolution equations and many third-order nonlinear evolution equations, such as KdV equation, Burgers equation. They also proposed a new residual neural network to solve the sine-Gordon equation\cite{LJCY2}. Wang and Yan\cite{WangL1} study the data-driven solutions of the defocused nonlinear NLS equation by using PINNs. Marcucci G et al\cite{MarcucciG} studied theoretically artificial neural networks with a nonlinear wave as a reservoir layer and developed a new computing model driven by nonlinear partial differential equations.

At present, deep learning method is only used to solve low-order nonlinear problems or low-order linear problems,and its applicability to high-order nonlinear problems is undiscovered. We apply the deep learning method and physics-informed neural networks\cite{MRaissi-2017} to solve high-order nonlinear equation and show the efficiency and effectiveness. Specifically, we will study the numerical driven solutions of the fourth-order Boussinesq equation and the fifth-order KdV equation.

Boussinesq equation, as a kind of nonlinear equation closely related to wave phenomena, has been widely studied in many fields of physics\cite{JBoussinesq}\cite{UrsellF}\cite{LuChangna}\cite{GuoB}\cite{HimonasAA}. The classical Boussinesq equation describes the evolution of waves in shallow water\cite{JBoussinesq}. Some progress has been made in the analytical solution of Boussinesq equation\cite{MaYL}\cite{HIETARINTAJ1}\cite{ZhangL}\cite{ZhangL}\cite{Clarkson PA}.

KdV equation has always been an important part of nonlinear mathematical and physical model, and it is also a hotspot of numerical methods\cite{KUOPY}\cite{VLIEGENTHARTAC}\cite{ZhangYingnan}. The numerical solution of the fifth-order KdV equation has also made some progress\cite{HuWeiPeng}\cite{AhmadHijaz}\cite{KayaD}.The numerical driven solution of KdV equation has been solved by Raissi M et al\cite{MRaissi-2017} and Li, Chen\cite{LJCY1}, Raissi M et al used discrete time model to solve KdV equation, Li and Chen used continuous time model to solve KdV equation. However, due to the complexity of high-order nonlinear equation, the numerical driven solution of high-order nonlinear problem have not been solved. We will illustrate the applicability of the deep learning method to the fifth-order nonlinear soliton equation by solving the fifth-order KdV equation.

The paper is organized as follows. In section \ref{sec:method intruction}, we will introduce the method of deep learning to solve high-order equations. In section \ref{sec:Boussinesq}, we use the deep learning method to reproduce the one-soliton and the two-soliton numerical driven solution of the fourth-order Boussinesq equation. In numerical driven solution, we find the dynamic behavior of solitons interaction. Specifically, we solve the chasing-soliton and collising-soliton of Boussinesq equation, and find the dynamic behavior between solitons from the numerical driven solution. In section \ref{sec:fifth-order KdV}, we get the one-soliton numerical driven solution and the two-soliton numerical driven solution of the fifth-order KdV equation. In the numerical driven solution, we also find the dynamic behavior of solitons interaction. Finally, some concluding discussions and remarks are contained in section \ref{sec:summary}.

\section{Method}\label{sec:method intruction}
We consider the following form of (1 + 1)-dimensional fourth-order and fifth-order nonlinear soliton equation,
\begin{equation}\label{pde-01}
\begin{aligned}
\chi_1(u_t,u_{tt})=\chi_2(u,u_{x},u_{xx},u_{xxx},u_{xxxx},u_{xxxxx}),
\end{aligned}
\end{equation}
and solve their soliton solution, where the subscripts $t$ and $x$ denote the partial derivatives, and $\chi_1$ is a linear function of the time derivative of $u(x,t)$, $\chi_2$ is a nonlinear function of $u(x,t)$ and its partial derivative to space variable $x$. Specifically, we build a multi-layer neural network to approximate the potential solution, and use the automatic differentiation technique to obtain its derivatives in time and space\cite{LiuDC}.

The residual network is defined
\begin{equation}\label{pde-02}
\begin{aligned}
f=\chi_1(u_t,u_{tt})-\chi_2(u,u_{x},u_{xx},u_{xxx},u_{xxxx},u_{xxxxx}).
\end{aligned}
\end{equation}

The shared parameters of neural networks can be learned by minimizing the loss of mean square error,
\begin{equation}\label{pde-03}
\begin{aligned}
LOSS=LOSS_u+LOSS_f ,
\end{aligned}
\end{equation}
\begin{equation}\label{pde-04}
\begin{aligned}
LOSS_u=\frac{1}{N_u}\sum_{n=1}^{N_u}|u(t^i_u,x^i_u)-u^i|\ , \\
LOSS_f=\frac{1}{N_f}\sum_{n=1}^{N_f}|f(t^i_f,x^i_f)|\ .
\end{aligned}
\end{equation}

\begin{figure}[h]
\vspace{-0.2cm}
  \setlength{\abovecaptionskip}{-0.5cm}
  \subfigure{\includegraphics[scale=0.4]{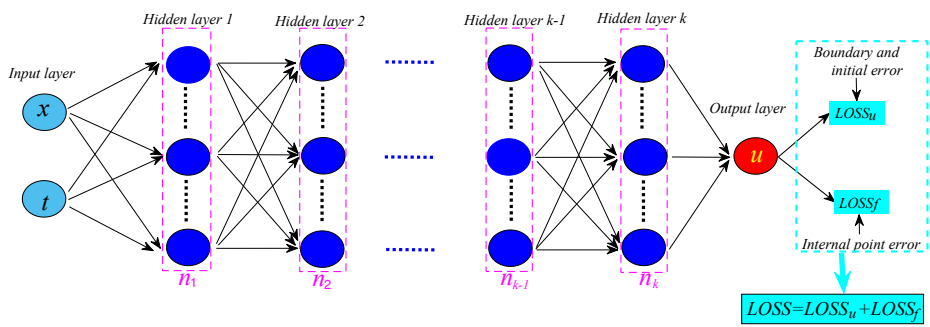}}
  \caption {Architecture of  physics-informed neural networks}
\label{fig:Neural network architecture}
\vspace{-0.2cm}
\end{figure}

$LOSS_u$ is the mean square error of initial and boundary, $LOSS_f$ is the internal calculation error, $t^i_u$ and $x^i_u$ represent the initial and boundary training values, $t^i_f$ and $x^i_f$ are the point collected in $f$. $N_u$ is the total number of selected boundary points and initial points, $N_f$ is the total number of selected internal collection points. A common deep feedforward neural fully connected network used to deal with high-order nonlinear problems. Figure \ref{fig:Neural network architecture} shows the framework of physics-informed neural networks. There are activation function, weights and bias between each layer. PINNs updates weights and biases by reducing error $LOSS$, and the neural network stops operation until the error is lower than the specified standard.

Where $k$ represent the number of hidden layers, $n_k$ represents the number of neurons corresponding to the hidden layer. We will improve the neural network from the following aspects to ensure that PINNs can deal with high-order nonlinear problems. Firstly, we will determine the appropriate number of hidden layers and the corresponding number of neurons. Secondly, we will improve the efficiency of neural network by improving the activation function.

In high-order nonlinear problem, it is very important to select the number of hidden layers and neurons in each layer. If the selected neural network structure is not suitable, there will occur over fitting( the deep learning model is superior in training set, but the final result is not good), under fitting( the deep learning model does not capture the characteristics of the data well and can not solve the equation well), loss of operational efficiency and other problems easily. Compared with low-order problems, high-order nonlinear problems are more difficult due to the complexity of their equation. Through a large number of experiments, we decide to use a neural network with four hidden layers to deal with high-order nonlinear soliton equation.

On the selection of activation function, Chen and Li\cite{LJCY1} proved that trigonometric function as activation function is effective on solitary wave solution of third-order nonlinear soliton equation. In this paper, we will study the deep learning algorithm of high-order nonlinear soliton equation and explore the effectiveness of trigonometric function as activation function for high-order nonlinear soliton equation and find the most suitable activation function. In addition, we use L-BFGS optimization algorithm\cite{LiuDC} to set all parameters of the target to minimize the loss function Equation (\ref{pde-03}). All numerical examples reported here are run on a Dell computer with Intel Xeon Gold 6320R i5 processor and 32 GB memory.

\section{Boussinesq equation}\label{sec:Boussinesq}

Boussinesq equation with Dirichlet boundary condition and initial condition given by
\begin{equation}\label{pde-05}
\left\{
\begin{array}{lr}
u_{tt}-u_{xx}-u_{xxxx}-3(u^2)_{xx}=0,x\in[-20,20],t\in[-5,5],
\\u(x,-5)=u_0(x),
\\u(-20,t)=u(20,t),
\end{array}
\right.
\end{equation}
where $u_0(x)$ is a given real valued smooth function.

We use the deep learning method to find the one-soliton and two-soliton numerical driven solution of equation (\ref{pde-05}) with tanh as activation function, and try to reproduce the dynamic behavior between solitons.
\subsection{One-soliton solution}\label{sec:Boussinesq1}
In this subsection, the numerical driven one-soliton solution of Boussinesq equation will be solved. The one-soliton analytic solution of Boussinesq equation can be obtained by using Hirota method\cite{MaYL}\cite{HIETARINTAJ1},
\begin{equation}\label{pde-b-1-u}
\begin{aligned}
u(x,t)=\frac{k_1^2}{2}sech^2\Bigg(\frac{k_1x+\sqrt{k_1^2+k_1^4}t }{2}+\xi_0\Bigg).
\end{aligned}
\end{equation}
We could set $k_1$ = 1, $\xi_0=0$. The corresponding initial condition becomes
\begin{equation}\label{pde-b-1-u0}
\begin{aligned}
u_0(x)=\frac{1}{2}sech^2\Bigg(\frac{x-5\sqrt{2} }{2}\Bigg).
\end{aligned}
\end{equation}

\begin{figure}[h]
\centering
  \subfigure[One-soliton Evolution of Exact Solution]{\includegraphics[scale=0.14]{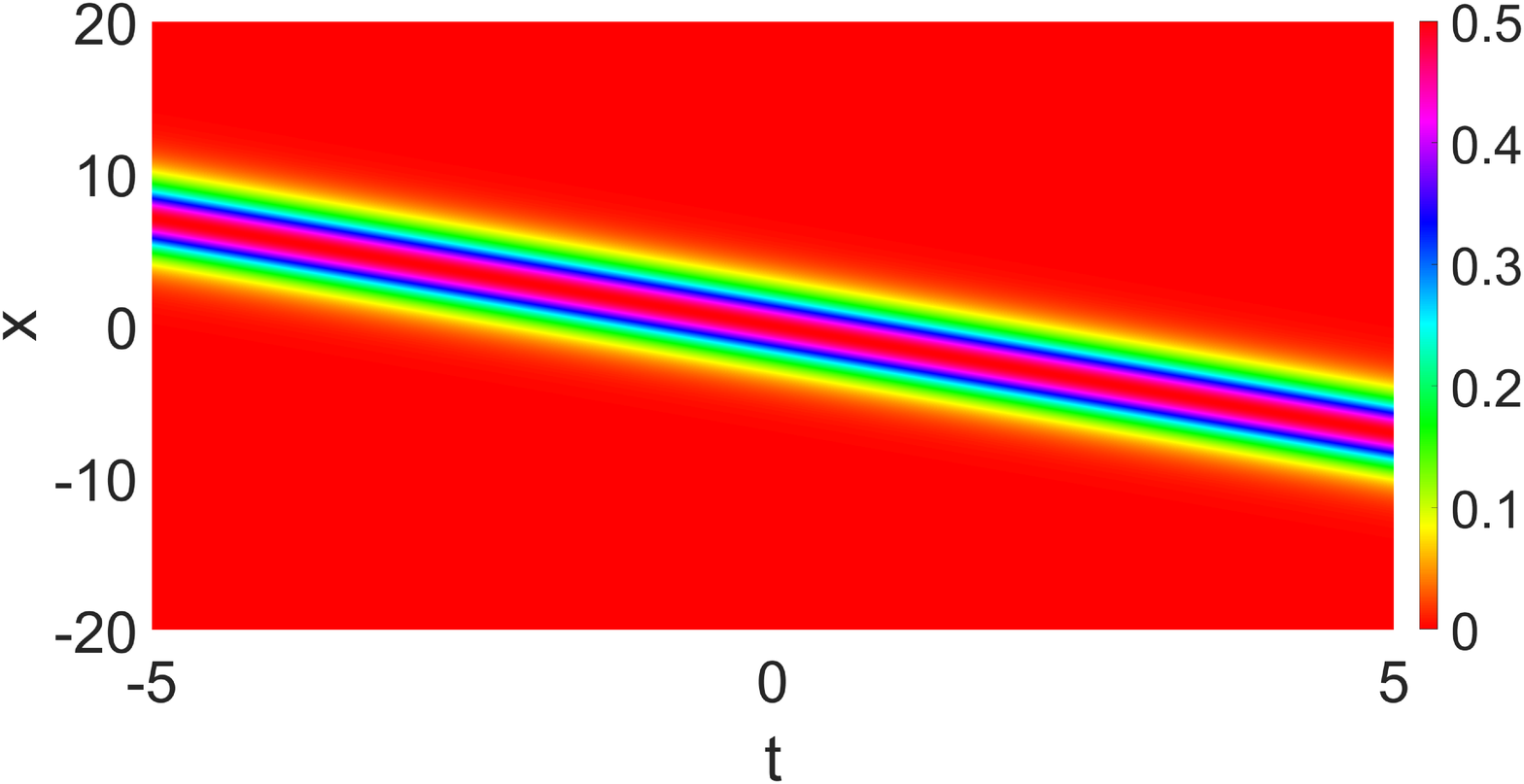}}
  \subfigure[One-soliton Evolution of Learned Solution] {\includegraphics[scale=0.14]{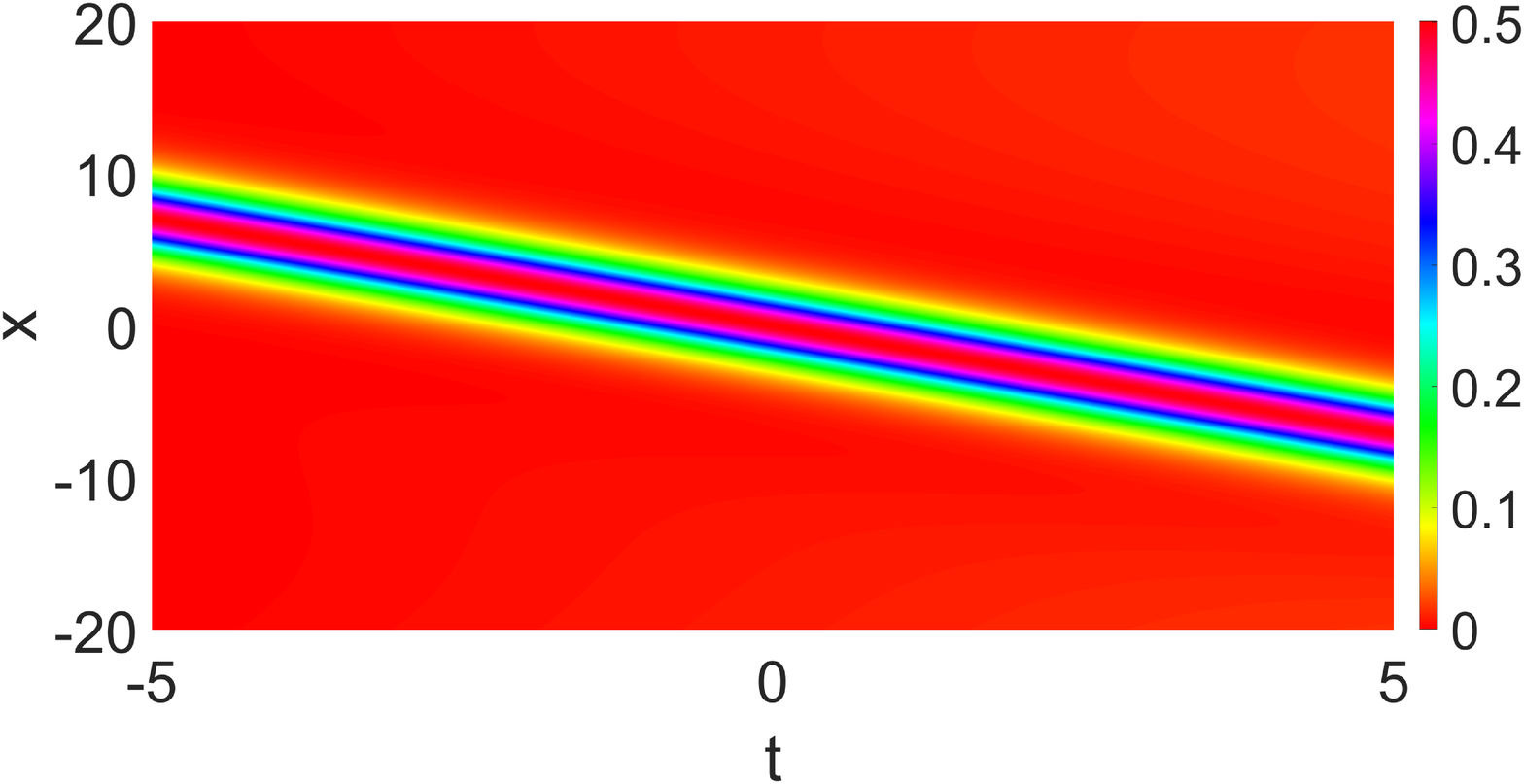}}\\
  \subfigure[$t=-2.5$]{\includegraphics[scale=0.3]{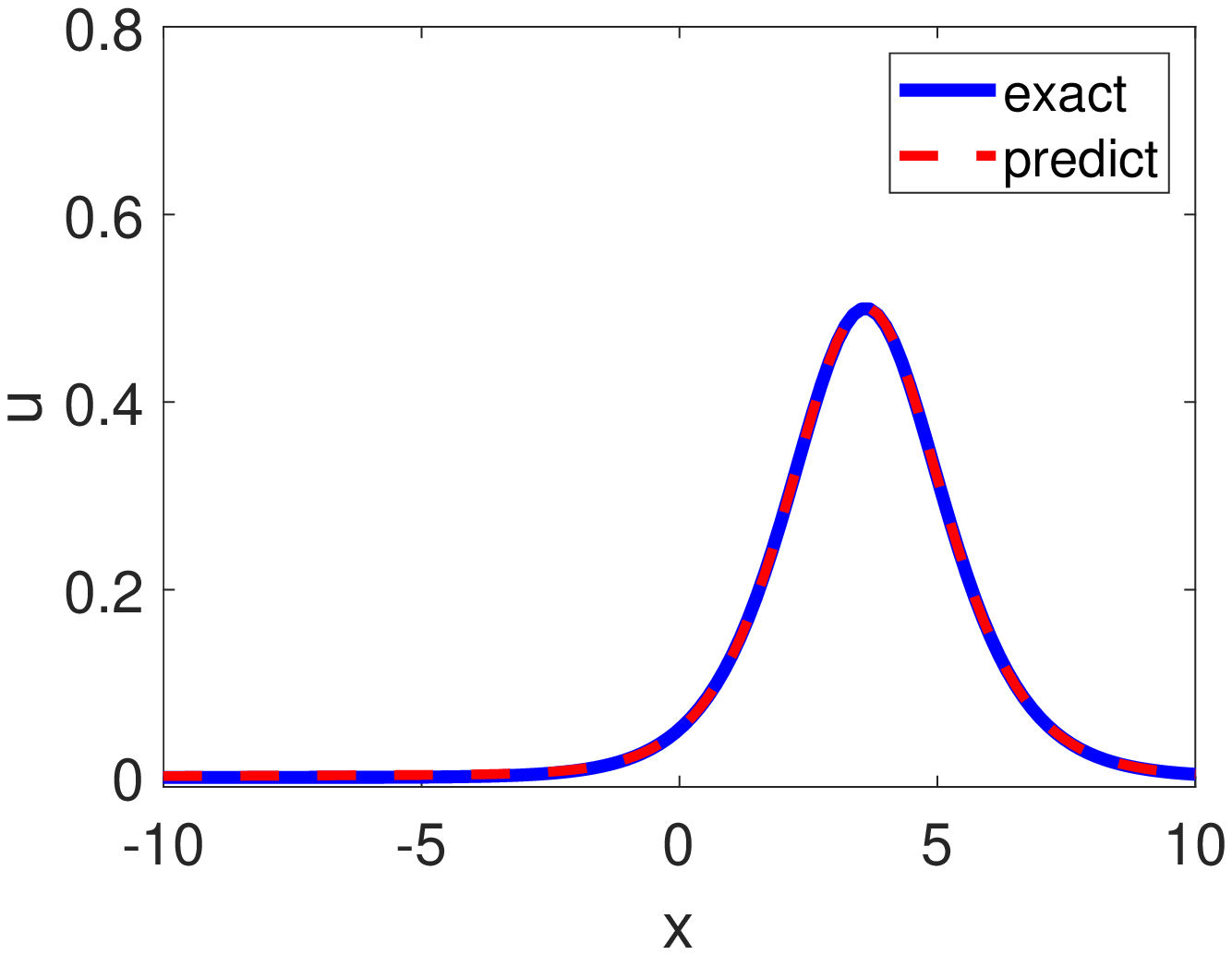}}
  \subfigure[$t=0$]{\includegraphics[scale=0.3]{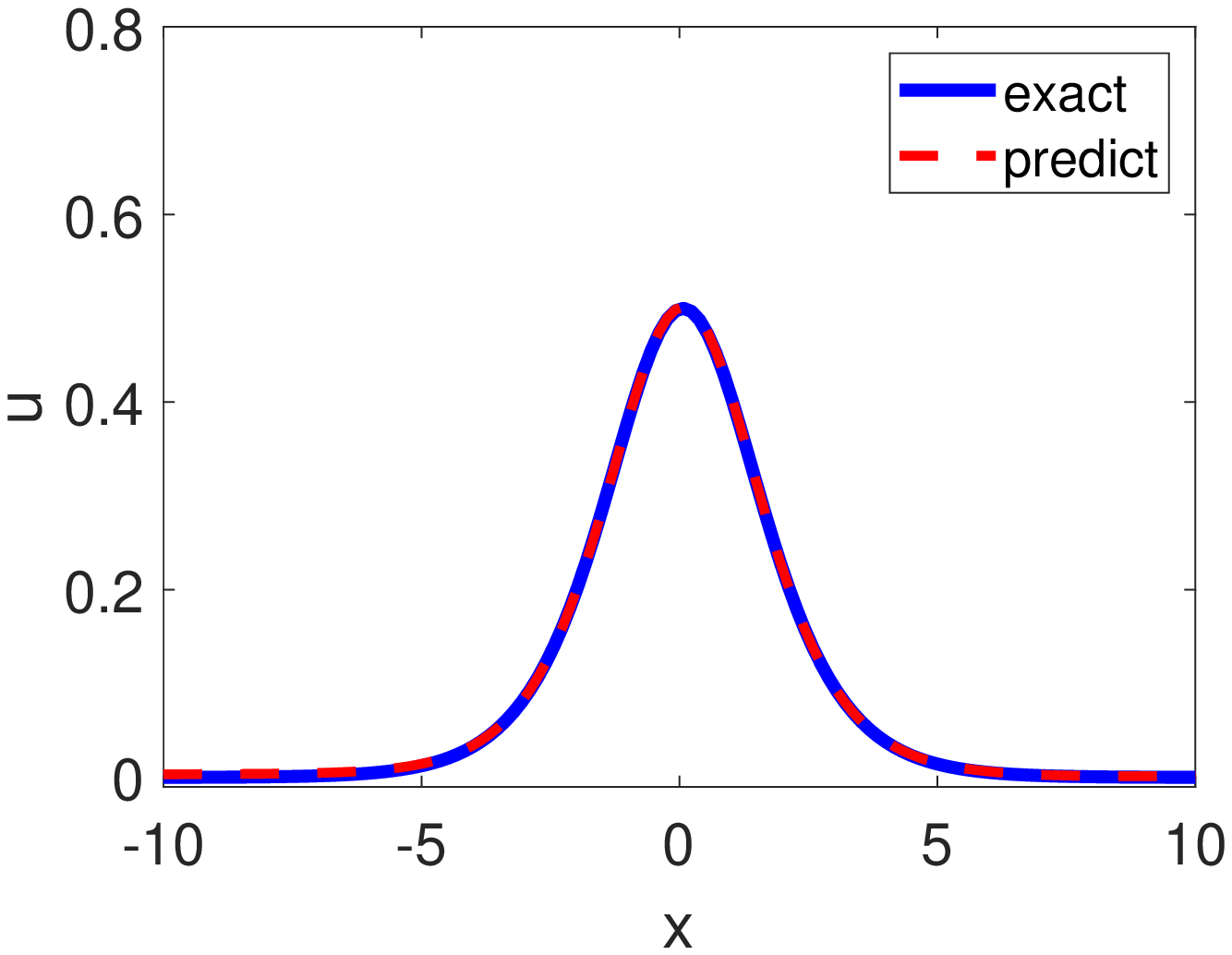}}
  \subfigure[$t=2.5$]{\includegraphics[scale=0.3]{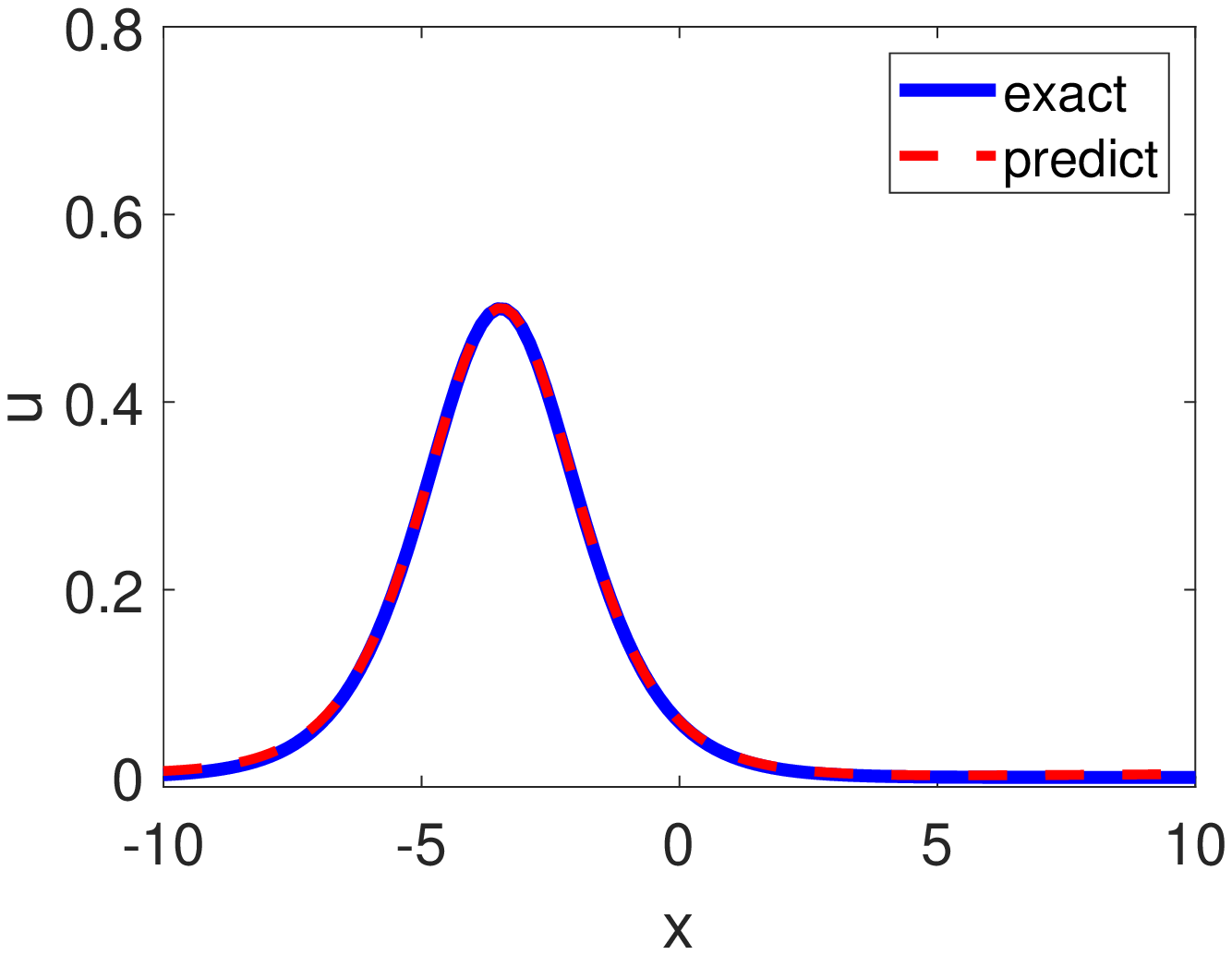}}
  \caption {Subgraph (a) and (b) are comparison of one-soliton exact solution and learned solution of Boussinesq equation, subgraph (c)-(e) are the detailed comparison of exact solution and learned spatiotemporal solution at the specific time.}
\label{fig:b-1-soliton}
\end{figure}

\begin{figure}
\centering
  \subfigure[One-soliton Numerical Driven Solution]{\includegraphics[scale=0.35]{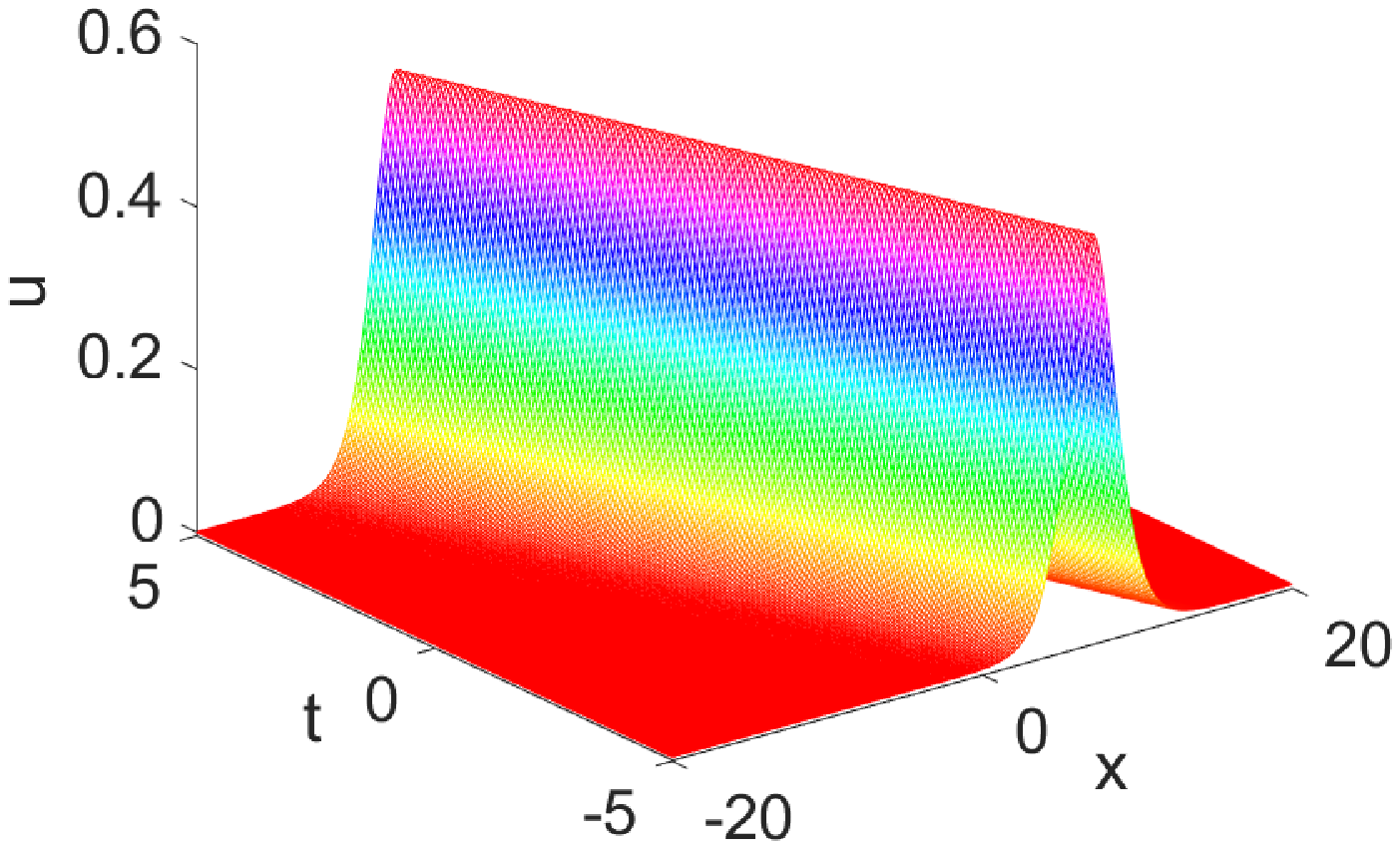}}
  \subfigure[One-soliton Exact Solution]{\includegraphics[scale=0.35]{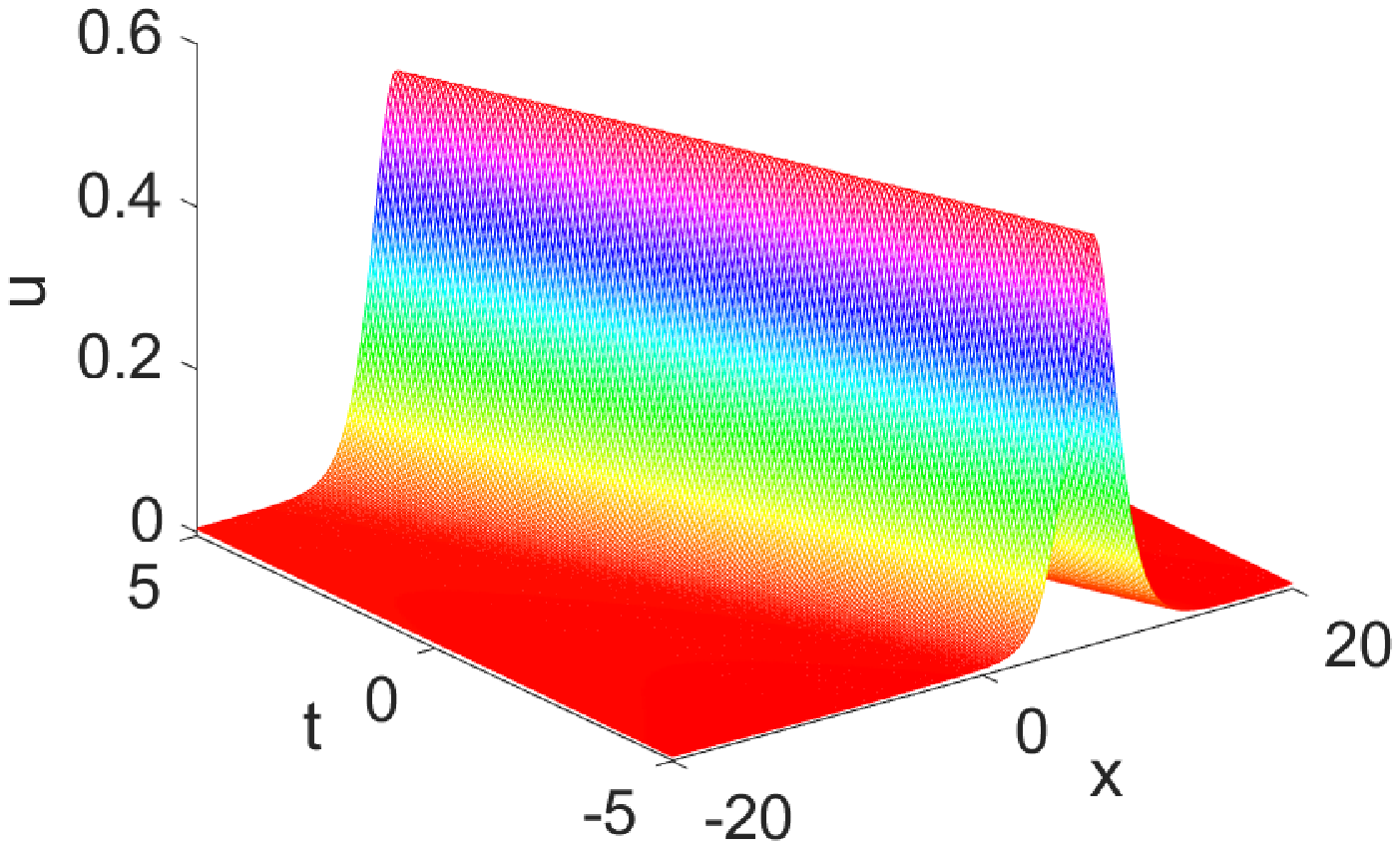}}
  \caption {Spatiotemporal evolution of one-soliton numerical driven solution and exact solution of Boussinesq equation.}
\label{fig:b-1-Spatiotemporal evolution}
\end{figure}

We generate the data of 201 snapshots directly on regular space-time grid with $\triangle t$ = 0.05s. A small training data subset is generated by randomly latin hypercube sampling method\cite{SteinML}, the number of collection points are $N_u$ = 100, $N_f$ = 20000. The latent solution $u(t,x)$ can be learned by minimizing the loss function Equation (\ref{pde-03}). Top panel of Figure \ref{fig:b-1-soliton} shows comparison of the predicted spatiotemporal solution and exact solution. The model achieves a relative $L^2$ error of size $2.0 \times 10^{-2}$ in a runtime of 152s. The model is iterated 170 times to complete the operation. Bottom panel of Figure \ref{fig:b-1-soliton} shows the detailed comparison of exact solution and predicted spatiotemporal solution at different times t = -2.5, t = 0, t = 2.5 respectively. One-soliton solution of Boussinesq equation is reconstructed by using deep learning method accurately. From Figure \ref{fig:b-1-Spatiotemporal evolution}, we can clearly observe the reconstructed solitary wave motion.

In order to verify the universality of our neural network architecture for the one-soliton solution of Boussinesq equation, we try to change the value of $k_1$ and give the numerical solution respectively. The results shows that our neural network architecture is effective in solving one-soliton of Boussinesq equation.
\begin{table}[H]\footnotesize
\vspace{-0.5cm}
\centering
\caption{The results of different one-soliton solution of Boussinesq equation calculated by deep learning method.}
\begin{tabular}{|c| c| c| c |c| c| c| c| c| }
  \hline
  $k_1$              & 0.8                & 0.9                & 1.0               & 1.1                & 1.2                 & 1.3                 & 1.4\\
  \hline
  $L^2$ error        &$2.79\times10^{-2}$  &$3.36\times10^{-2}$  &$2.03\times10^{-2}$ & $1.9\times10^{-2}$ & $1.67\times10^{-2}$ & $1.60\times10^{-2}$ & $1.62\times10^{-2}$ \\
  \hline
  Time(s)            &224                 &191                 &152                & 160                & 294                 & 284                 & 480  \\
  \hline
  Iterations         &141                 &155                 &170                & 195                & 576                 & 416                 & 753  \\
  \hline
\end{tabular}\label{tab:boussinesq1-k1}
\end{table}

\subsection{ Two-soliton solution}\label{sec:Boussinesq2}

In this subsection, we will calculate the numerical driven two-soliton solution of the Boussinesq equation and reproduce the solitons interaction process. The two-soliton solution of Boussinesq equation is given\cite{MaYL}\cite{HIETARINTAJ1}.
\begin{equation}\label{pde-07}
\centering
\begin{aligned}
u(x,t)=2\frac{k_1^2e^{k_1x+\omega_1t+\delta_1}+k_2^2e^{k_2x+\omega_2t+\delta_2}+(k_1+k_2)^2e^{(k_1+k_2)x+(\omega_1+\omega_2)t+\delta_1+\delta_2+\delta_0}}{1+e^{k_1x+\omega_1t+\delta_1}+e^{k_2x+\omega_2t+\delta_2}+e^{(k_1+k_2)x+(\omega_1+\omega_2)t+\delta_1+\delta_2+\delta_0}} \\-2\frac{(k_1e^{k_1x+\omega_1t+\delta_1}+k_2e^{k_2x+\omega_2t+\delta_2}+(k_1+k_2)e^{(k_1+k_2)x+(\omega_1+\omega_2)t+\delta_1+\delta_2+\delta_0})^2}{(1+e^{k_1x+\omega_1t+\delta_1}+e^{k_2x+\omega_2t+\delta_2}+e^{(k_1+k_2)x+(\omega_1+\omega_2)t+\delta_1+\delta_2+\delta_0})^2}, \\
e^{\delta_0}=- \frac{(\omega_1-\omega_2)^2-(k_1-k_2)^2-(k_1-k_2)^4 }{(\omega_1+\omega_2)^2-(k_1+k_2)^2-(k_1+k_2)^4}, \\
\end{aligned}
\end{equation}
where $\delta_1$ and $\delta_2$ is constant, $\omega_1$and $\omega_2$ meet the conditions $\omega_1^2=k_1^2+k_1^4$ , $\omega_2^2=k_2^2+k_2^4$ respectively.

Two-soliton solution of Boussinesq equation have two states, colliding-soliton( the two solitons have different directions) and chasing-soliton( the two solitons have the same direction and different amplitude). We will find the numerical driven solution of the two forms respectively, and study their dynamic behavior and the interaction between solitons.
\subsubsection{Colliding-soliton}
\begin{figure}[ht]
\vspace{-0.2cm}
\centering
  \setlength{\abovecaptionskip}{0.cm}
  \setlength{\belowcaptionskip}{-0.cm}
  \subfigure[Colliding-soliton Interaction of Exact Solution]{\includegraphics[scale=0.14]{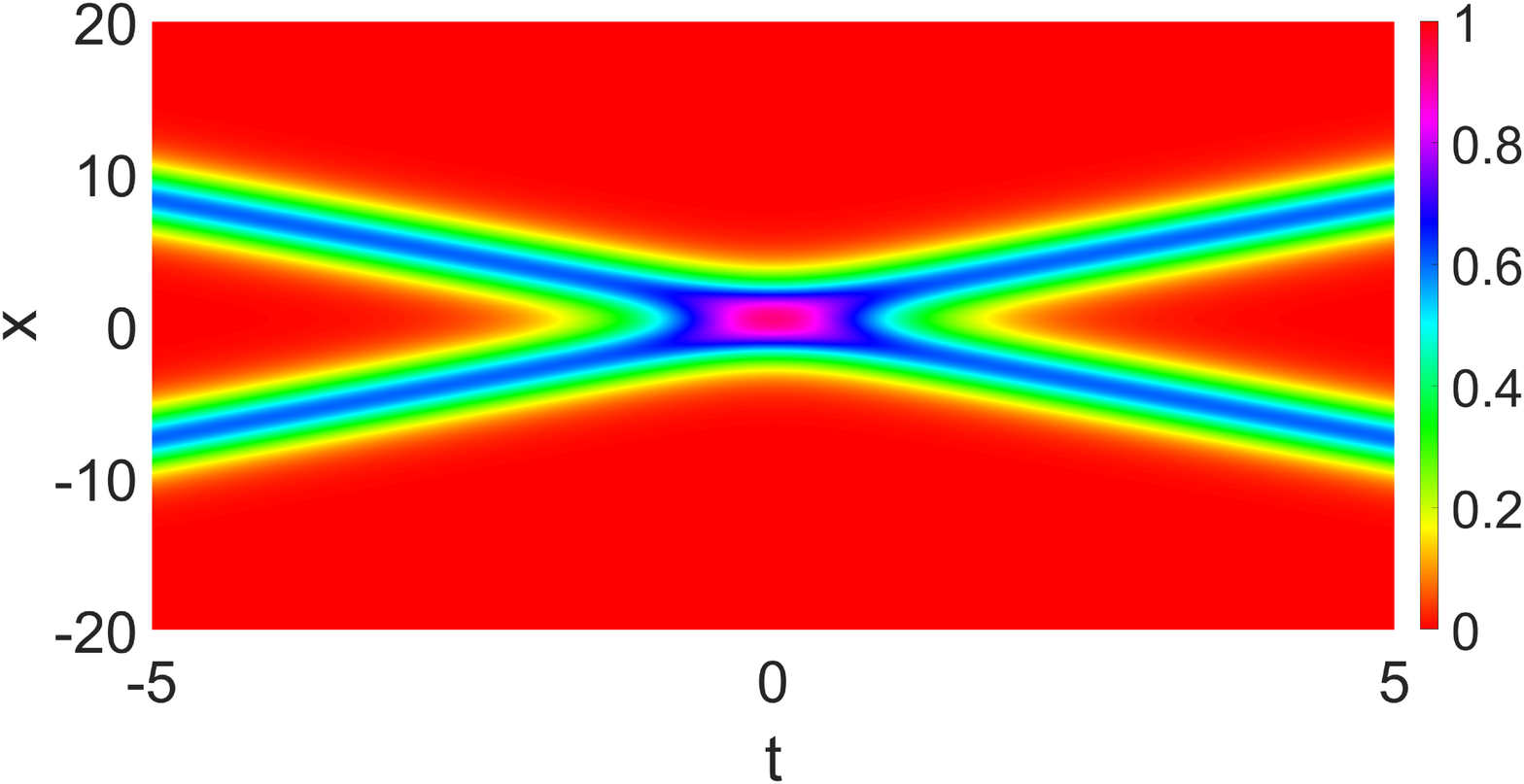}}
  \subfigure[Colliding-soliton Interaction of Learned Solution] {\includegraphics[scale=0.14]{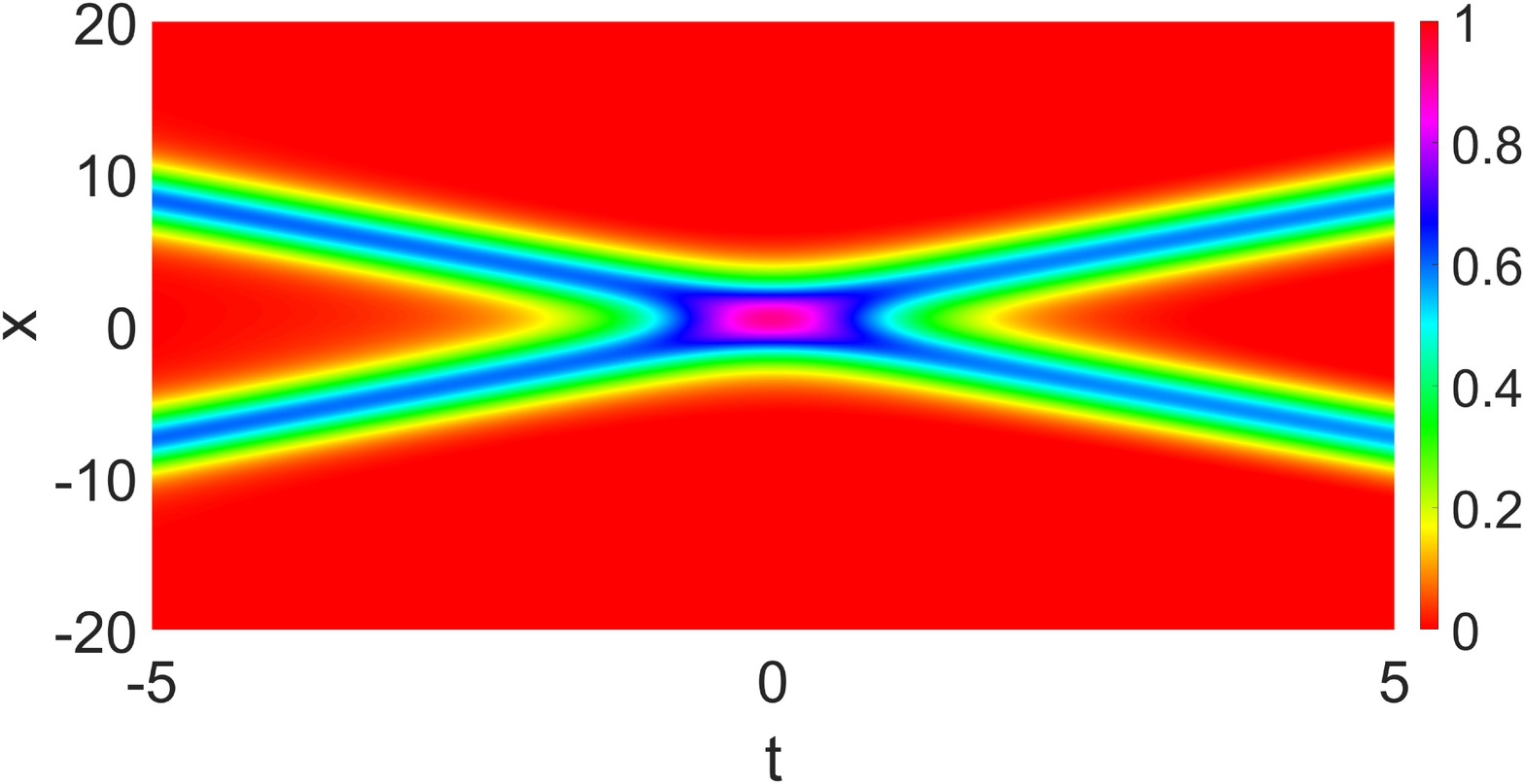}}\\
  \subfigure[$t=-4.5$]{\includegraphics[scale=0.3]{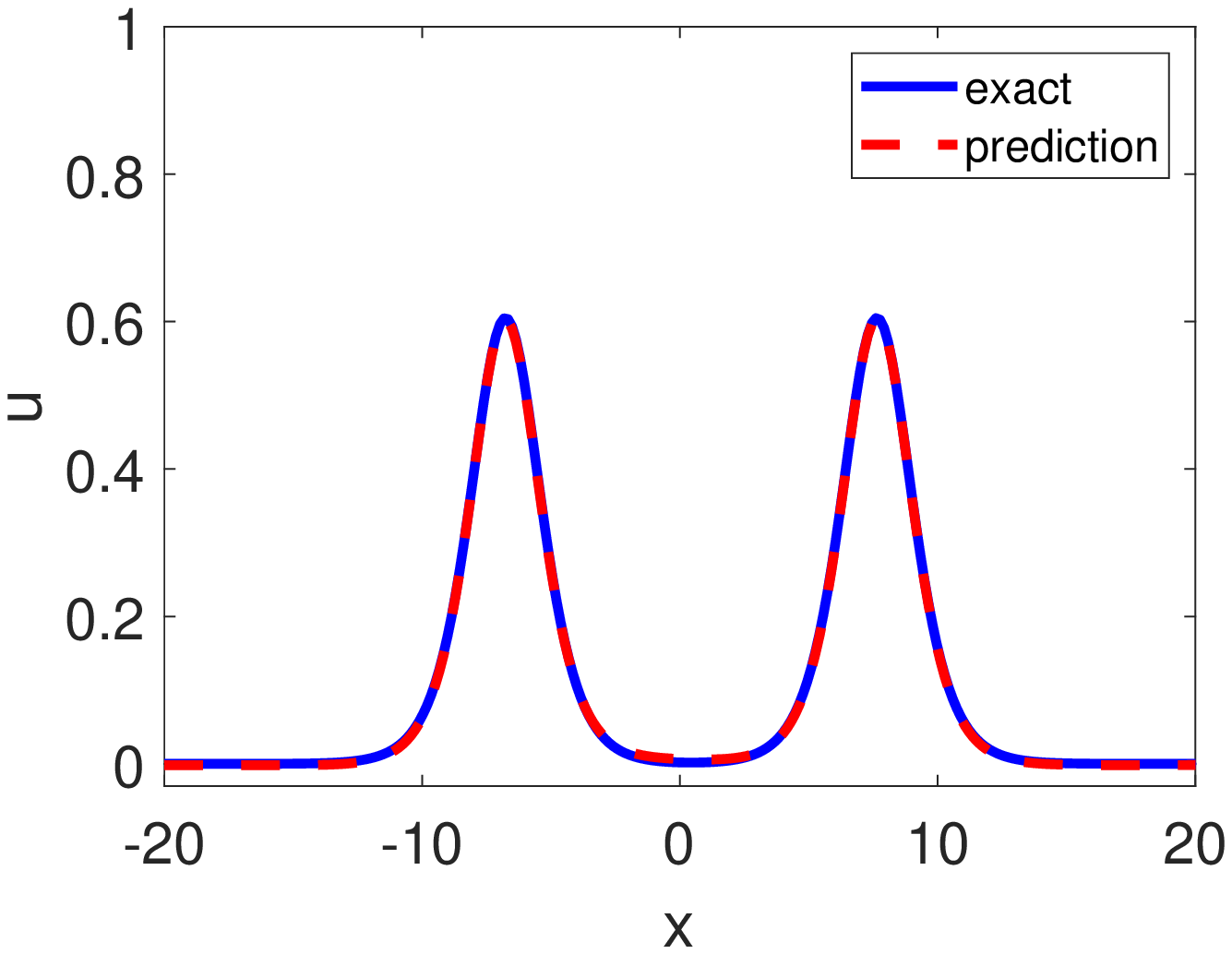}}
  \subfigure[$t=0$]{\includegraphics[scale=0.3]{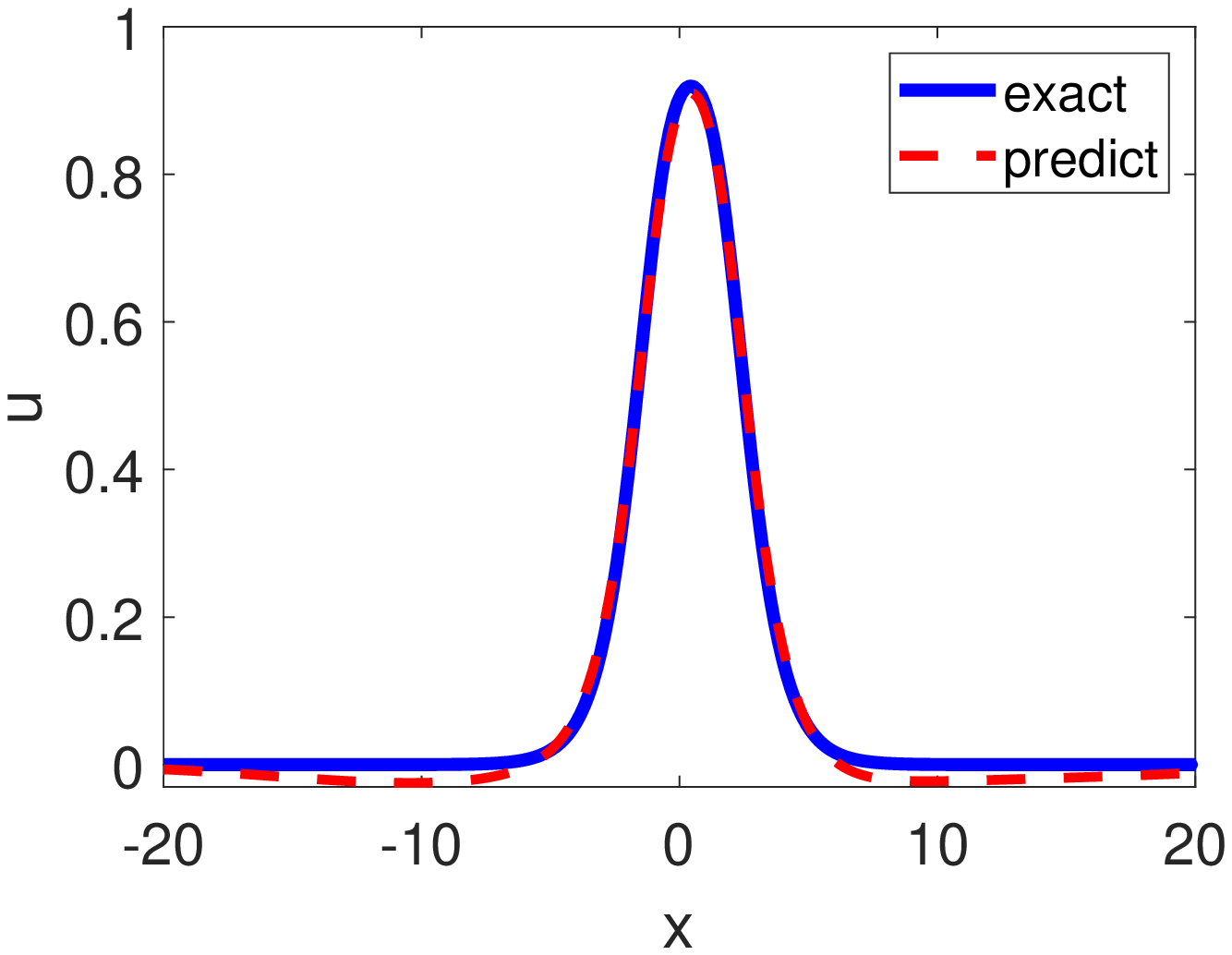}}
  \subfigure[$t=2.5$]{\includegraphics[scale=0.3]{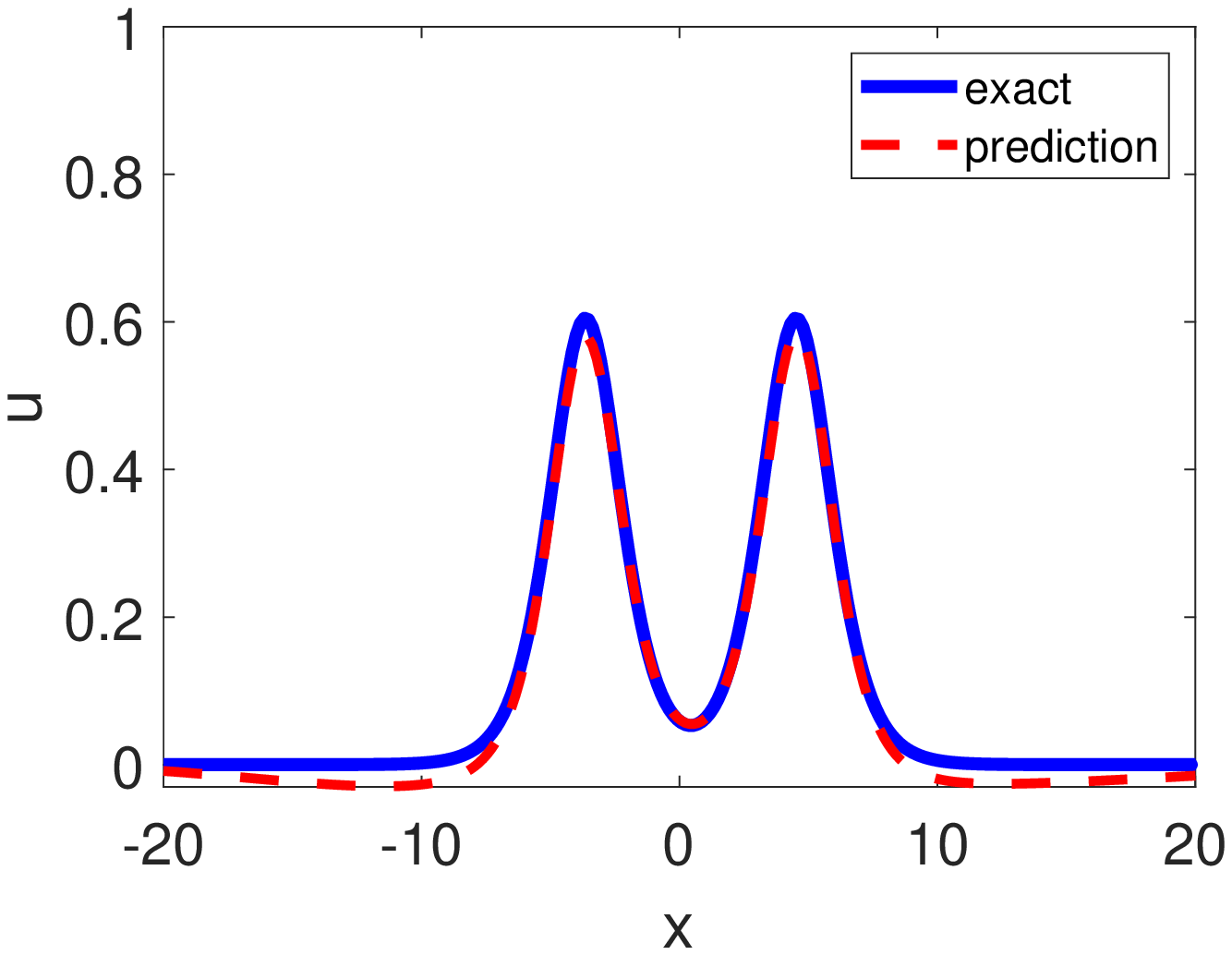}} \\
  \caption {Subgraph (a) and (b) are comparison of colliding-soliton exact solution and learned solution of Boussinesq equation, subgraph (c)-(e) are the detailed comparison of exact solution and learned spatiotemporal solution at the specific time.}
\label{fig:b-2-p-soliton}
\end{figure}

\begin{figure}[ht]
\centering
  \setlength{\abovecaptionskip}{0.cm}
  \setlength{\belowcaptionskip}{-0.cm}
  \subfigure[Colliding-soliton Numerical Driven Solution]{\includegraphics[scale=0.35]{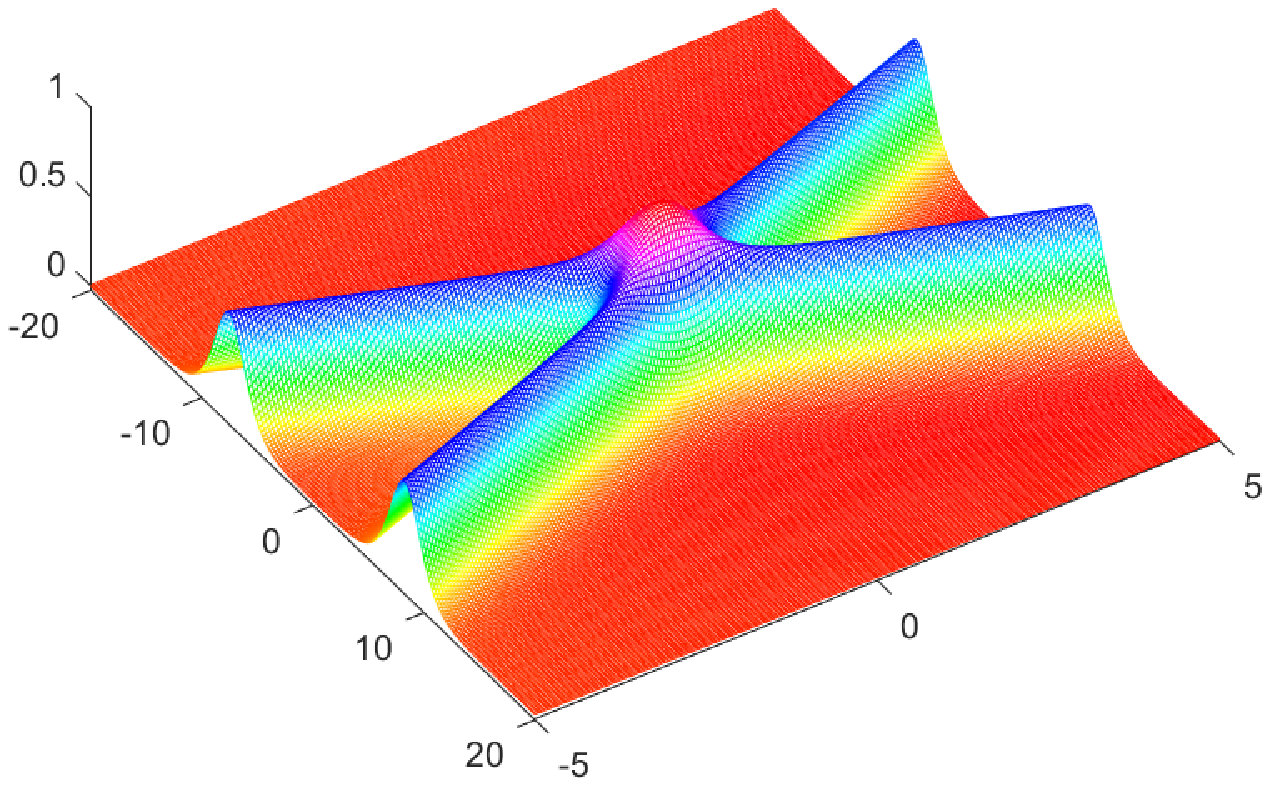}}
  \subfigure[Colliding-soliton Exact Solution]{\includegraphics[scale=0.35]{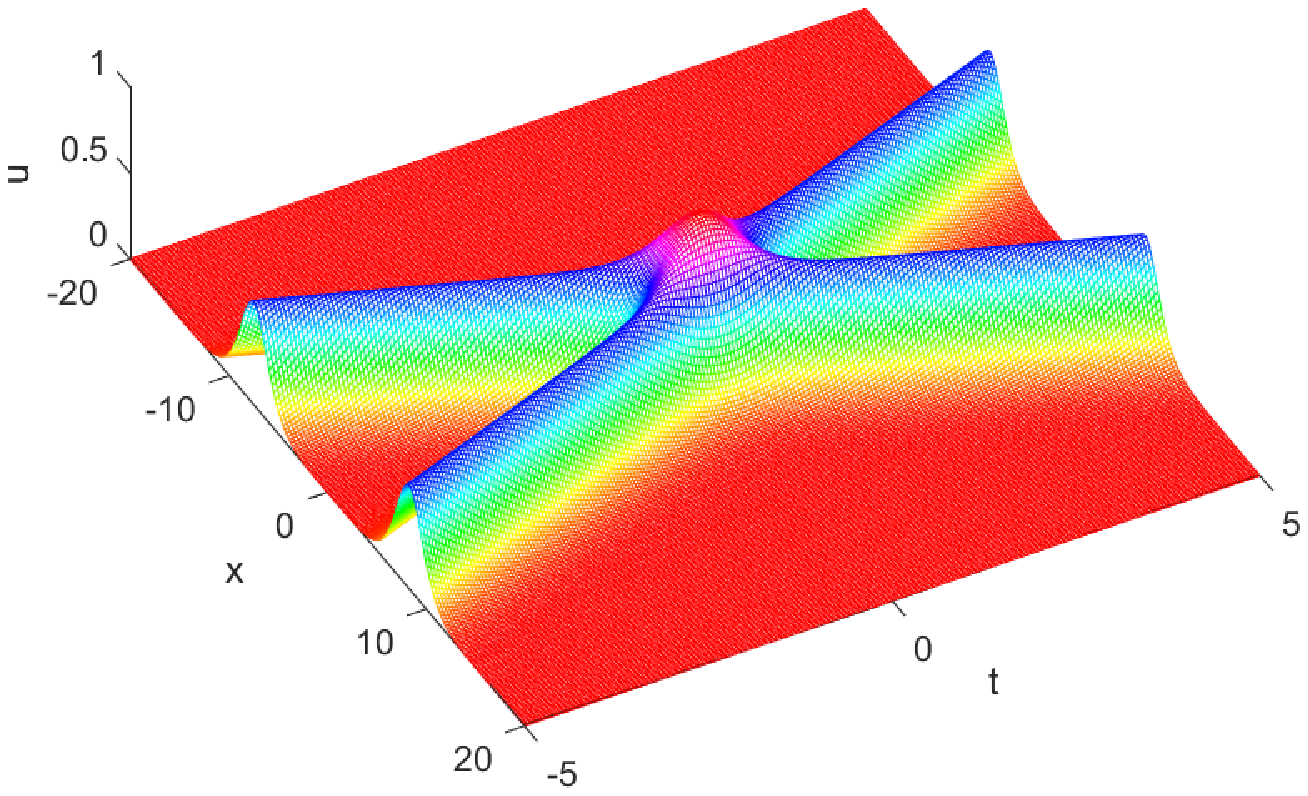}}
  \caption {Spatiotemporal evolution of colliding-soliton numerical driven solution and exact solution.}
\label{fig:b-2-p-Spatiotemporal evolution}
\end{figure}

We just set $k_1$ = $k_2$ = 1.1, $\delta_1$ = $\delta_2$ =0, $\omega_1$and $\omega_2$ have opposite sign. We generate the data of 201 snapshots directly on regular space-time grid with $\triangle t$ = 0.05s. A small training data subset is generated by randomly latin hypercube sampling method\cite{SteinML}, the number of collection points are $N_u$ = 100, $N_f$ = 25000. The latent solution $u(x,t)$ is learned by minimizing the loss function Equation (\ref{pde-03}). Top panel of Figure \ref{fig:b-2-p-soliton} shows the comparison of predicted spatiotemporal solution and exact solution. Bottom panel of Figure \ref{fig:b-2-p-soliton} shows the detailed comparison of actual solution and predicted spatiotemporal solution at different time t = -4.5, t = 0, t = 2.5 respectively. The specific spatiotemporal evolution of colliding-soliton of Boussinesq equation is given in Figure \ref{fig:b-2-p-Spatiotemporal evolution}. The model achieves a relative $L^2$ error of size $7.63 \times 10^{-2}$ in a runtime of 430s. The model is iterated 578 times to complete the operation.

From Figure \ref{fig:b-2-p-Spatiotemporal evolution}, we learn the spatiotemporal evolution process of separation-fusion-separation of colliding-soliton. Amplitude becomes high during the colliding process and shape remains unchanged before and after the interaction, which is consistent with the known fact. The `phase shift' phenomenon also occur in numerical driven solution.

In order to verify the universality of our neural network architecture for the colliding-soliton, we calculate the different colliding-soliton solutions of Boussinesq equation by deep learning method. The result shows that the deep learning method is effective in solving the colliding-soliton solution of Boussinesq equation.
\begin{table}[H]\footnotesize
\vspace{-0.5cm}
\centering
\caption{The results of different colliding-soliton solution of Boussinesq equation  calculated by deep learning method.}
\begin{tabular}{|c| c| c| c |c| c| c| }

  \hline
  $k_1$                 & 0.8                & 0.9                & 1.0               & 1.1                & 1.2                                  \\
  \hline
  $k_2$                 & 0.8                & 0.9                & 1.0               & 1.1                & 1.2                                 \\
  \hline
  $L^2$ error         &$8.55\times10^{-2}$  &$8.10\times10^{-2}$  & $1.16\times10^{-1}$ & $7.63\times10^{-2}$ & $7.64\times10^{-2}$ \\
  \hline
  time(s)             &358                 &551                  & 642                   & 430                & 1065                            \\
  \hline
  Iterations          &320                 &635                  & 801                   & 578                & 755                           \\
  \hline
\end{tabular}\label{tab:boussinesq1-k2}
\end{table}

\subsubsection{Chasing-soliton}
\begin{figure}[h]
\vspace{-0.2cm}
\centering
\setlength{\abovecaptionskip}{-0.1cm}
\setlength{\belowcaptionskip}{-0.1cm}
  \subfigure[Chasing-soliton Interaction of Exact Solution]{\includegraphics[scale=0.14]{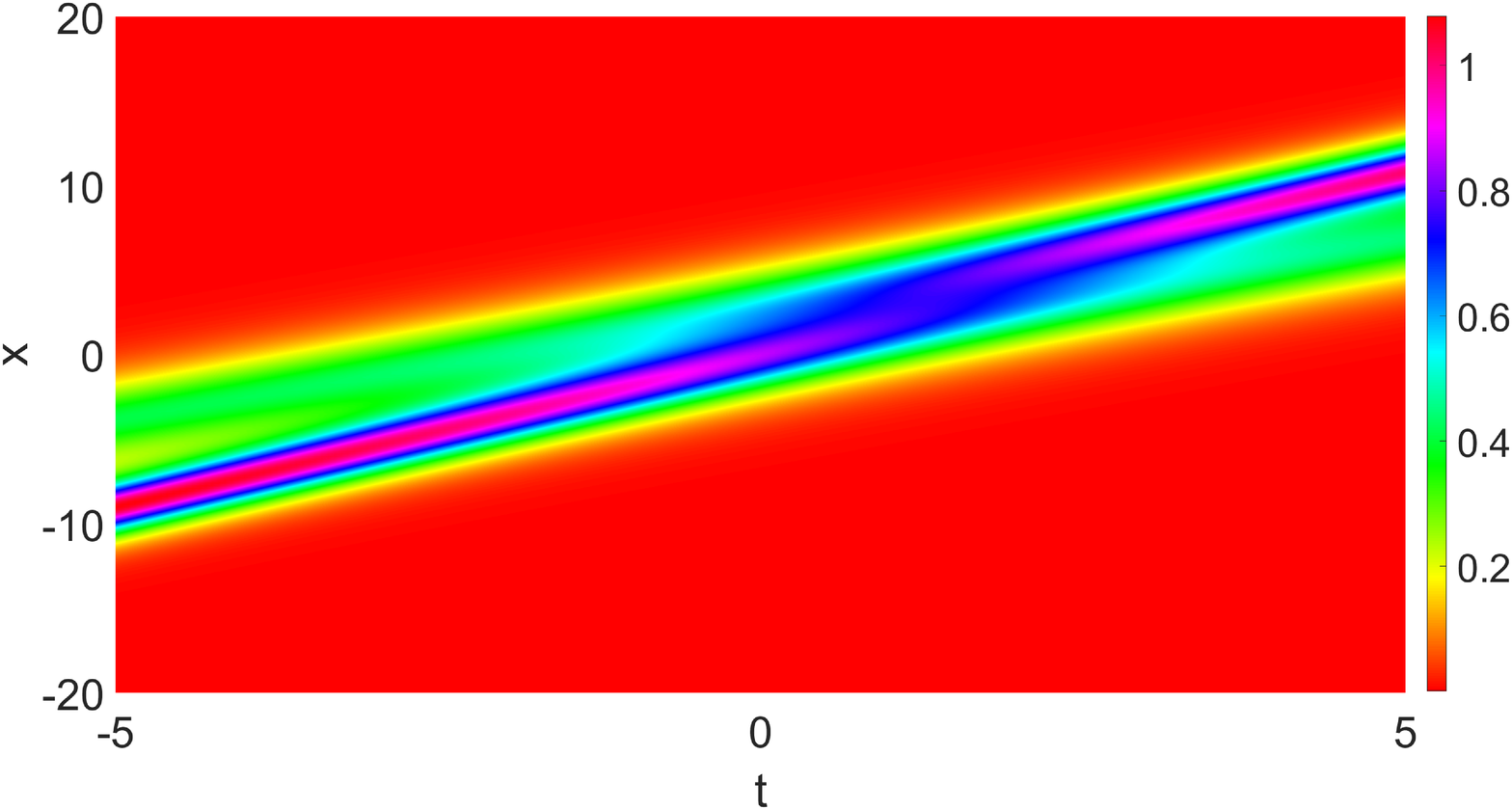}}
  \subfigure[Chasing-soliton Interaction of Learned Solution] {\includegraphics[scale=0.14]{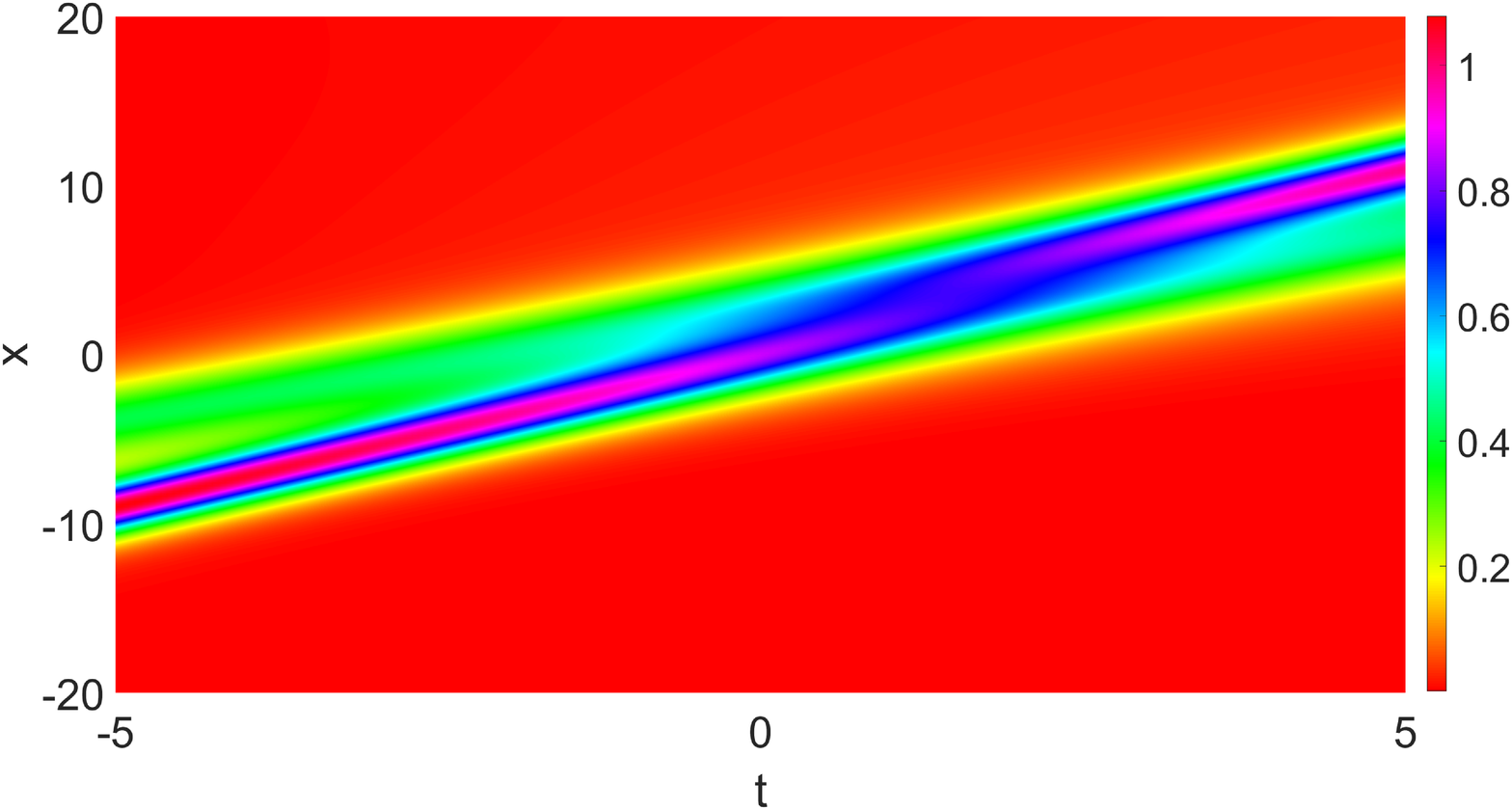}}\\
  \subfigure[$t=-4.5$]{\includegraphics[scale=0.3]{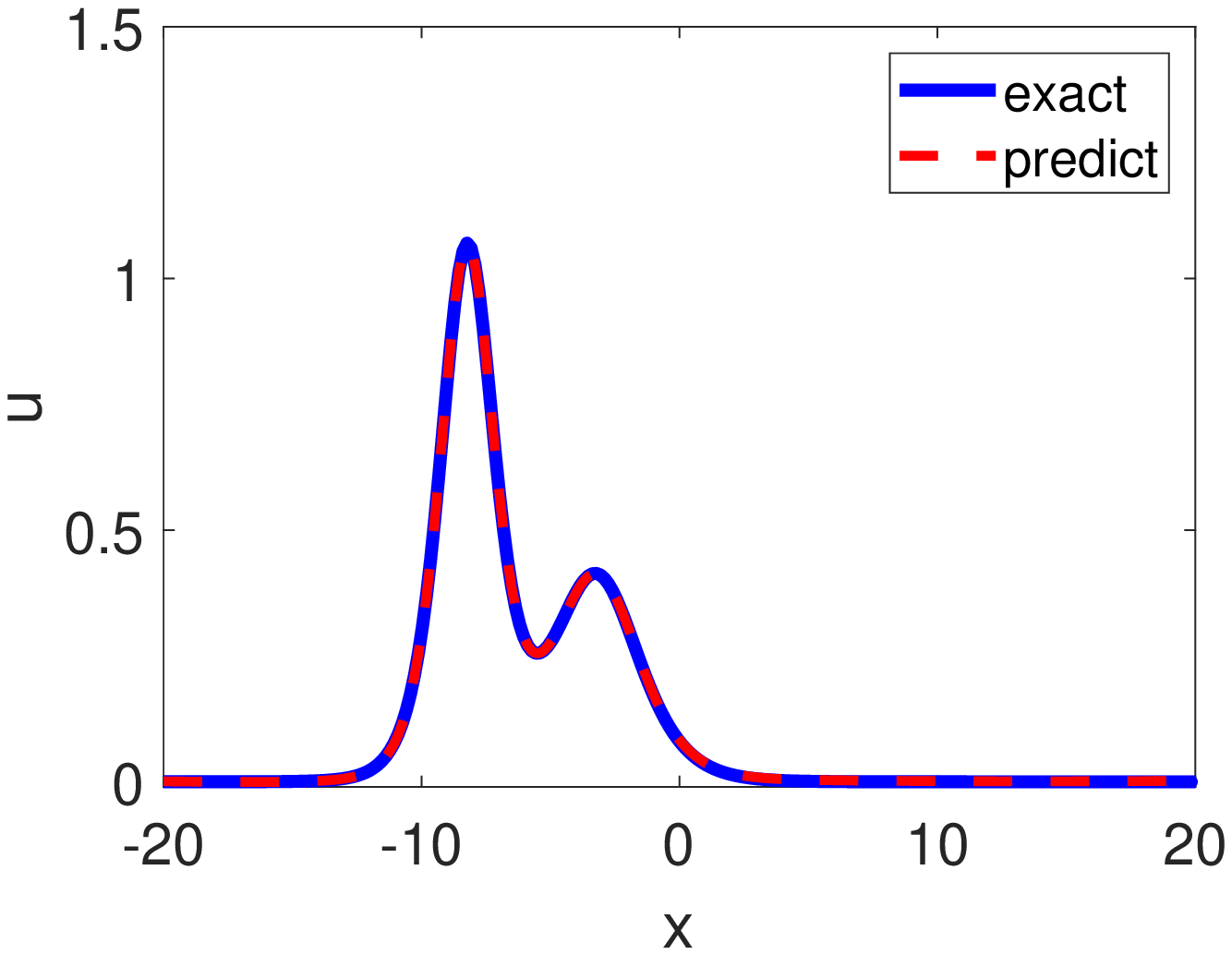}}
  \subfigure[$t=0$]{\includegraphics[scale=0.3]{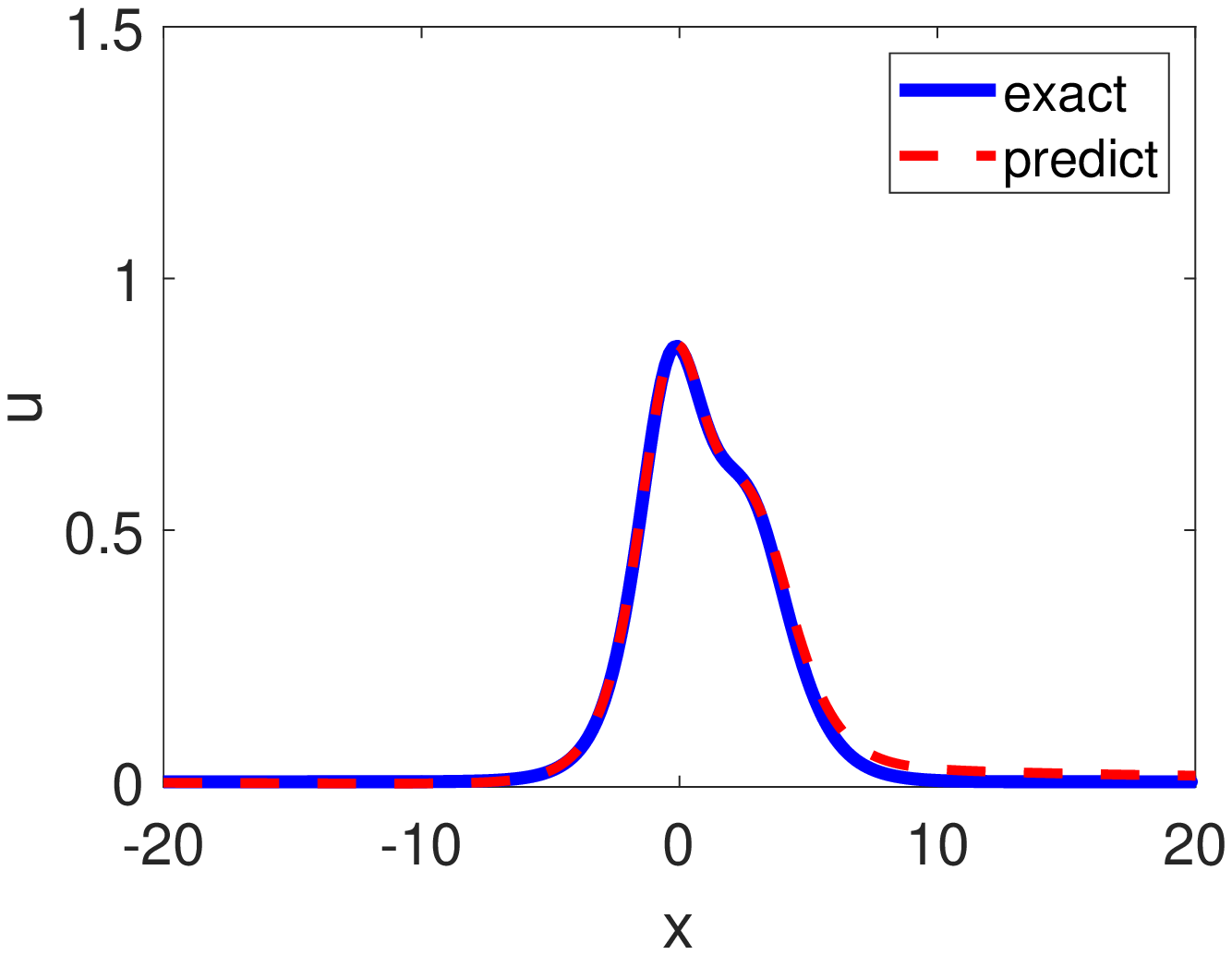}}
  \subfigure[$t=2.5$]{\includegraphics[scale=0.3]{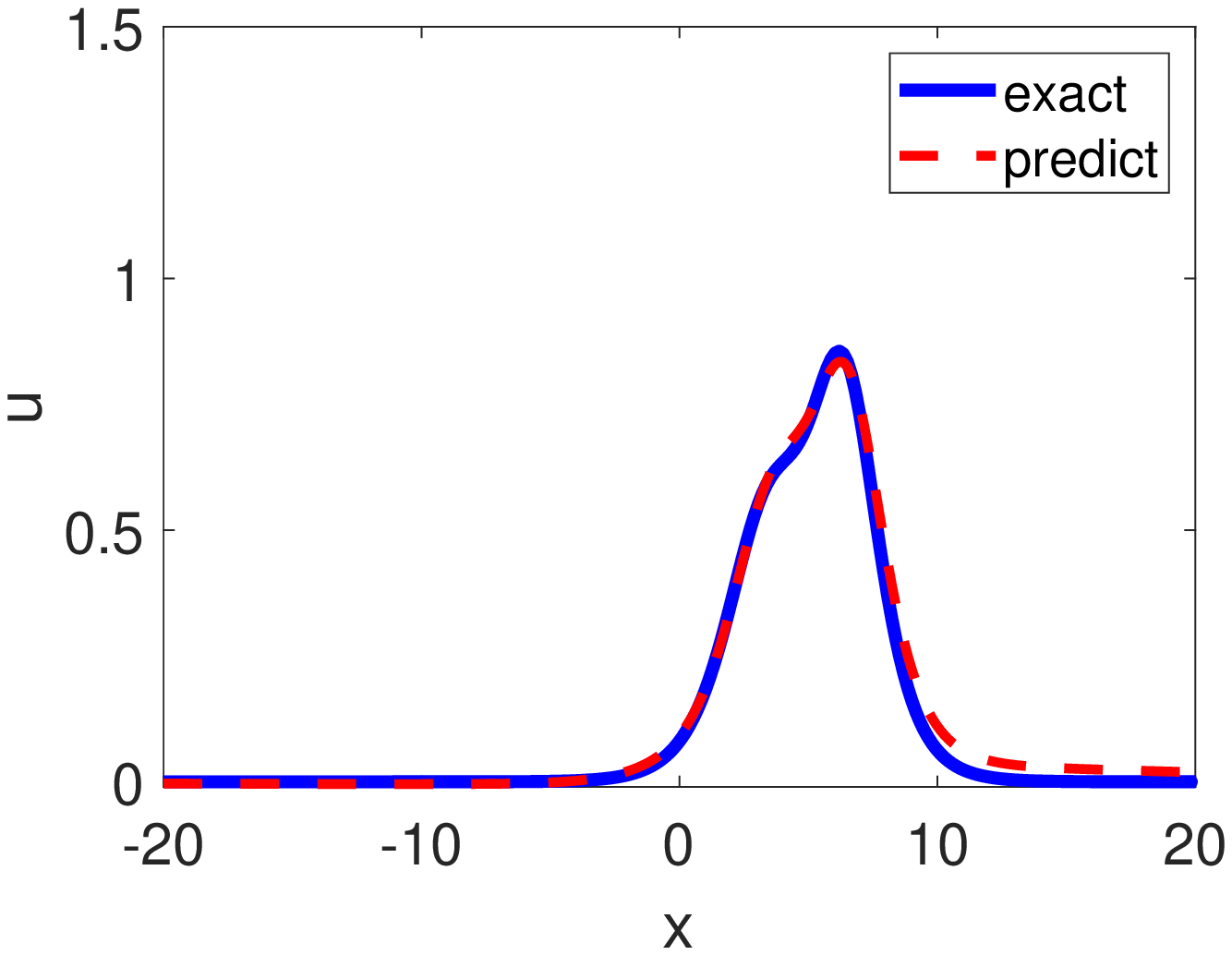}}
  \caption{Subgraph (a) and (b) are comparison of chasing-soliton exact solution and learned solution of Boussnesq equation, subgraph (c)-(e) are the detailed comparison of exact solution and learned spatiotemporal solution at the specific time.}
\label{fig:b-2-z-soliton}
\vspace{-0.3cm}
\end{figure}

We could set $k_1$ =1.5, $k_2$ = 0.9, $\delta_1$ = $\delta_2$ =0, $\omega_1$and $\omega_2$ are positive. In this condition, two-solitons have same direction and different magnitude, so the soliton chasing phenomenon occurs. We generate the data of 201 snapshots directly on the regular space-time grid with $\triangle t$ = 0.05s. A small training data subset is generated by randomly latin hypercube sampling method\cite{SteinML}, the number of collection points are $N_u$ = 100, $N_f$ = 25000. Top panel of \ref{fig:b-2-z-soliton} shows the comparison of predicted spatiotemporal solution and exact solution. Bottom panel of Figure \ref{fig:b-2-z-soliton} shows the detailed comparison of exact solution and learned spatiotemporal solution at different time  t = -4.5, t = 0, t = 2.5 respectively. The model achieves a relative $L^2$ error of size $6.73 \times 10^{-2}$ in a runtime of 892s.

\begin{figure}[h]
\vspace{-0.3cm}
\centering
  \subfigure[ Numerical Driven Solution]{\includegraphics[scale=0.35]{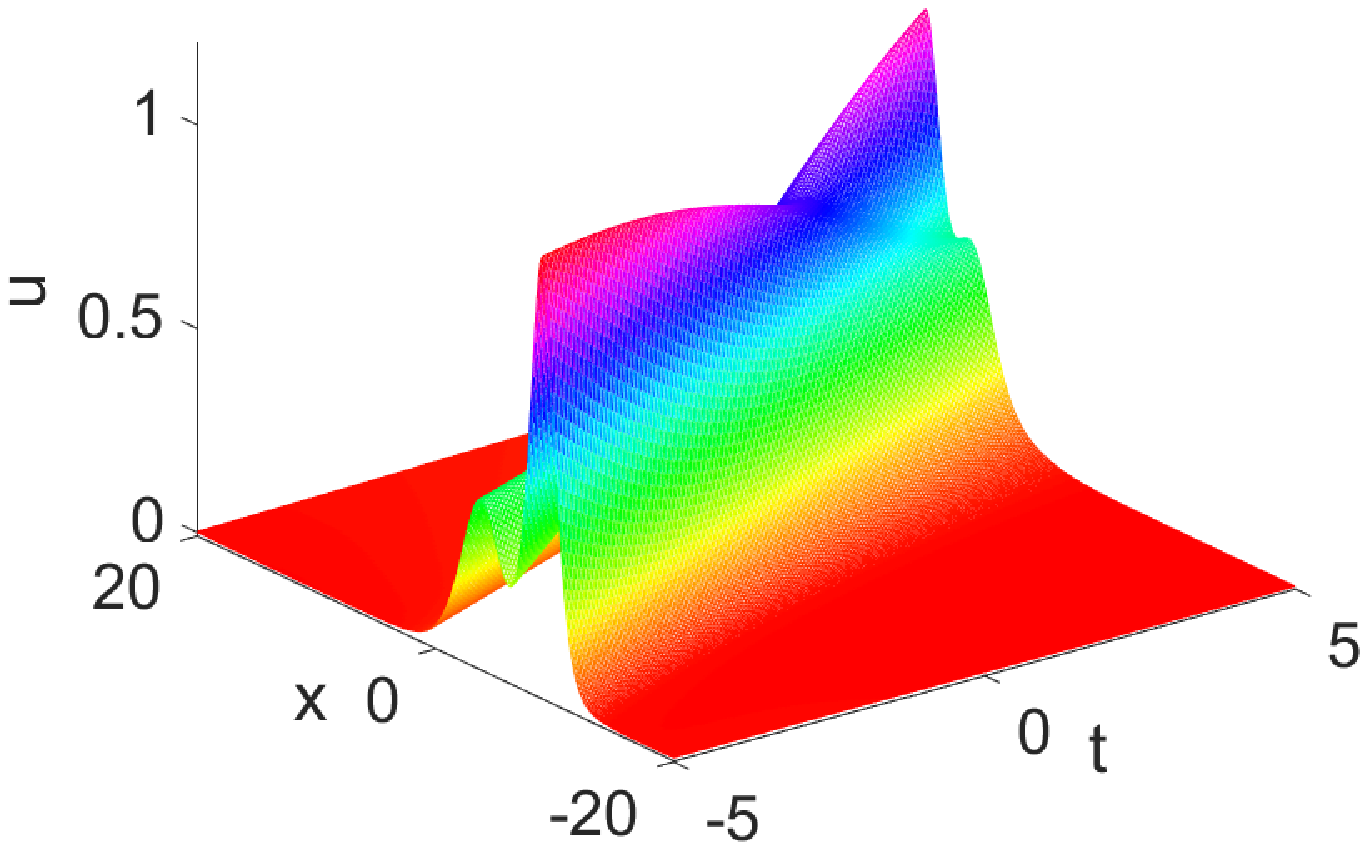}}
  \subfigure[Exact Solution]{\includegraphics[scale=0.35]{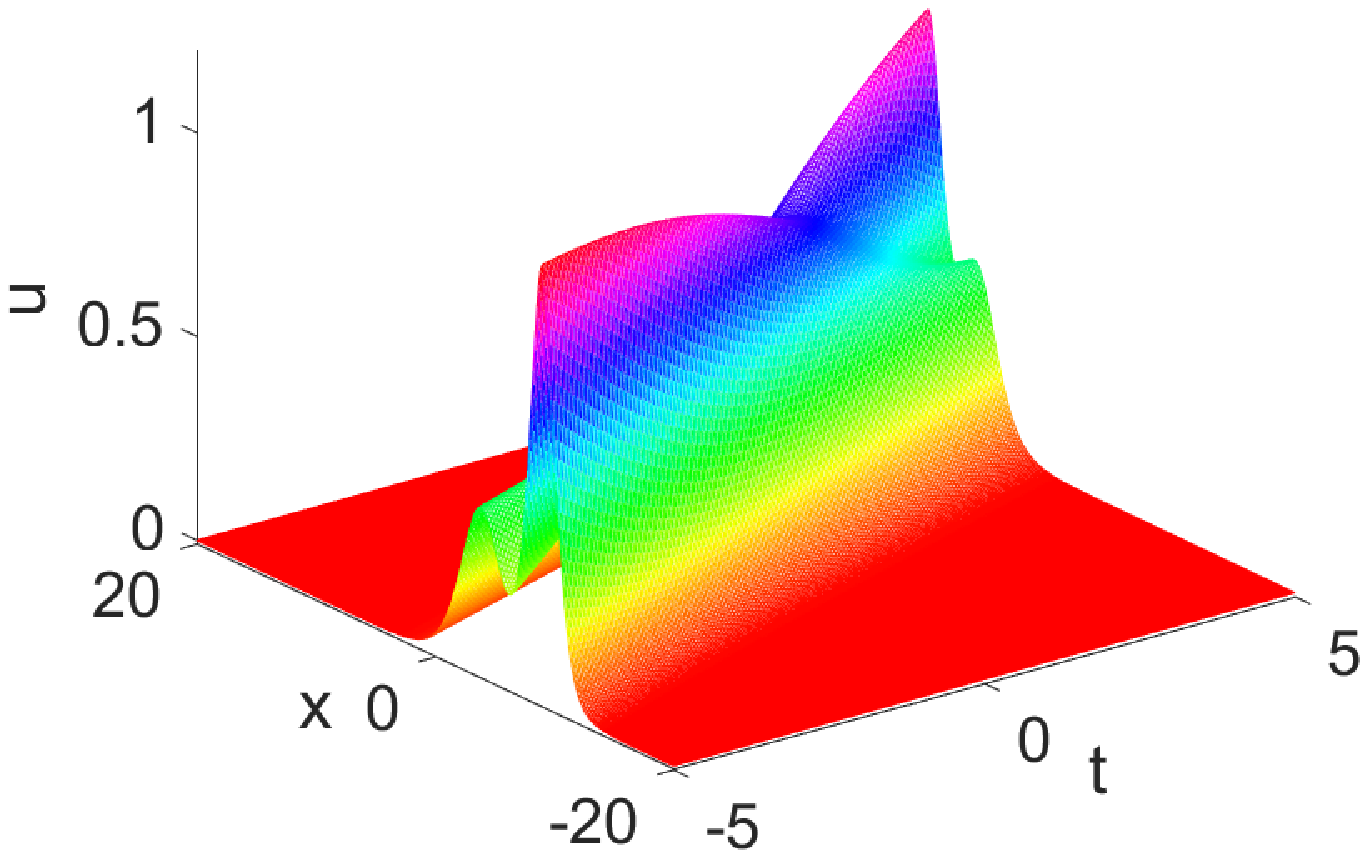}}
  \caption {Spatiotemporal evolution of chasing-soliton numerical driven solution and exact solution.}
\label{fig:b-2-z-spatiotemporal evolution}
\vspace{-0.2cm}
\end{figure}

Figure \ref{fig:b-2-z-spatiotemporal evolution} shows the spatiotemporal evolution process of separation-fusion-separation. Amplitude becomes low during the fusion process and shape remains unchanged before and after the interaction, which is consistent with the known fact. We also observe the `phase shift' phenomenon in the chasing-soliton numerical driven solution.

In exploring the effectiveness of activation function, we find tanh function is more effective than trigonometric function. We take one-soliton solution as an example, the calculation results is shown in Tabel \ref{tab:compare}. We find that both tanh and trigonometric function is useful in Boussinesq equation, and tanh cost less computational source. Compared with one-soliton solution, the two-soliton solution of Boussinesq equation cost more computational source.

\begin{table}[H]\footnotesize
\vspace{-0.5cm}
\centering
\caption{The results of different activation function of one-soliton calculated by deep learning method.}
\begin{tabular}{|c| c| c| c |c| c| c| }

  \hline
  activation function   & tanh              & cos                & sin               & sigmoid                & relu                                  \\
  \hline
  $L^2$ error         &$1.97\times10^{-2}$  &$1.86\times10^{-1}$  & $1.27\times10^{-1}$ & $9.22\times10^{-1}$  & $8.37\times10^{-1}$ \\
  \hline
  time(s)             &160                 &1670                  & 1531                   & 105                & 16                            \\
  \hline
  Iterations          &195                 &4722                  & 3080                   & 0                 & 4                           \\
  \hline
\end{tabular}\label{tab:compare}
\end{table}
\section{Fifth-order KdV equation}\label{sec:fifth-order KdV}

In this section, we consider the fifth-order KdV equation with Dirichlet periodicity boundary condition and initial condition,
\begin{equation}\label{fifth-kdv1}
\left\{
\begin{array}{lr}
u_{t}+(\alpha u_{xxxx}+\beta uu_{xx}+\gamma u^3)_x =0,x\in[-10,10],t\in[0,2\pi],
\\u(x,0)=u_0(x),
\\u(-10,t)=u(10,t),
\end{array}
\right.
\end{equation}
where $\alpha$, $\beta$, $\gamma$ are arbitrary constant, $u_0(x)$ is a given real valued smooth function.

We choose cos as activation function, and explore the effectiveness of trigonometric function as activation function of the fifth-order KdV equation. A deep learning method is used to find the one-soliton and two-soliton solution of the equation, and reproduce the dynamic behavior between the solitons.
\subsection{One-soliton solution}\label{sec:kdv1}
Using Hirota bilinear method, the one-soliton analytical solution of fifth-order KdV equation (\ref{fifth-kdv1}) can be obtained\cite{HIETARINTAJ2},
\begin{equation}\label{kdv-01}
\begin{aligned}
u(x,t)=\frac{15\alpha k_1^2}{2\beta}sech^2\Bigg(\frac{k_1x-{\alpha k_1^5}t +\xi_0}{2}\Bigg).
\end{aligned}
\end{equation}
We set $\alpha$=1, $\beta$=15, $\gamma$=15, $k_1$ = 1, $\xi_0=3$, the equation is also called C-D-J-K equation. Correspondingly,
\begin{equation}\label{pde-kdv-1-u0}
\begin{aligned}
u_0(x)=\frac{1}{2}sech^2\Bigg(\frac{x}{2}-1\Bigg).
\end{aligned}
\end{equation}

\begin{figure}[ht]
\centering
\setlength{\abovecaptionskip}{0.cm}
\setlength{\belowcaptionskip}{-0.cm}
  \subfigure[One-soliton Evolution of Exact Solution]{\includegraphics[scale=0.14]{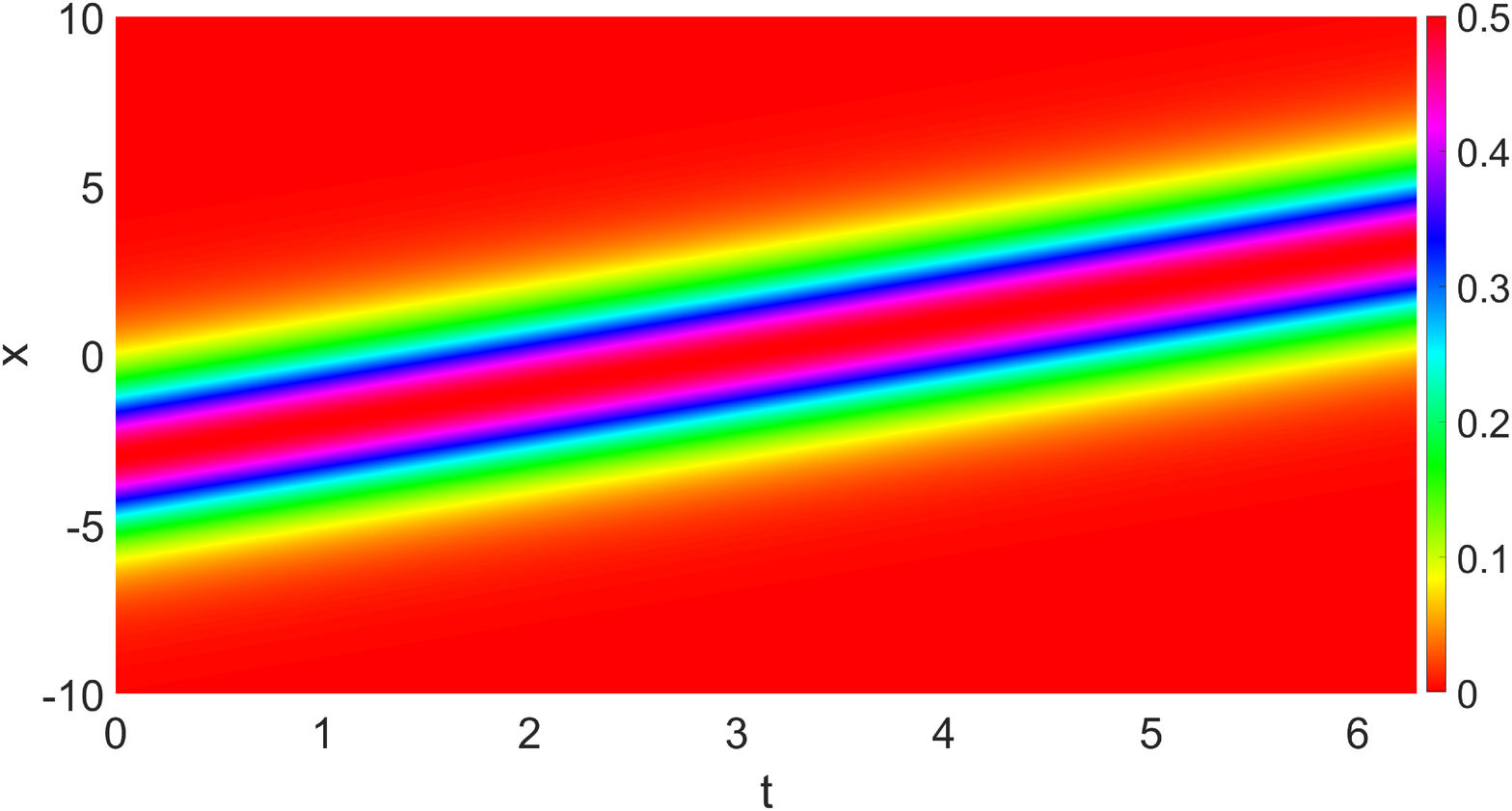}}
  \subfigure[One-soliton Evolution of Learned Solution] {\includegraphics[scale=0.14]{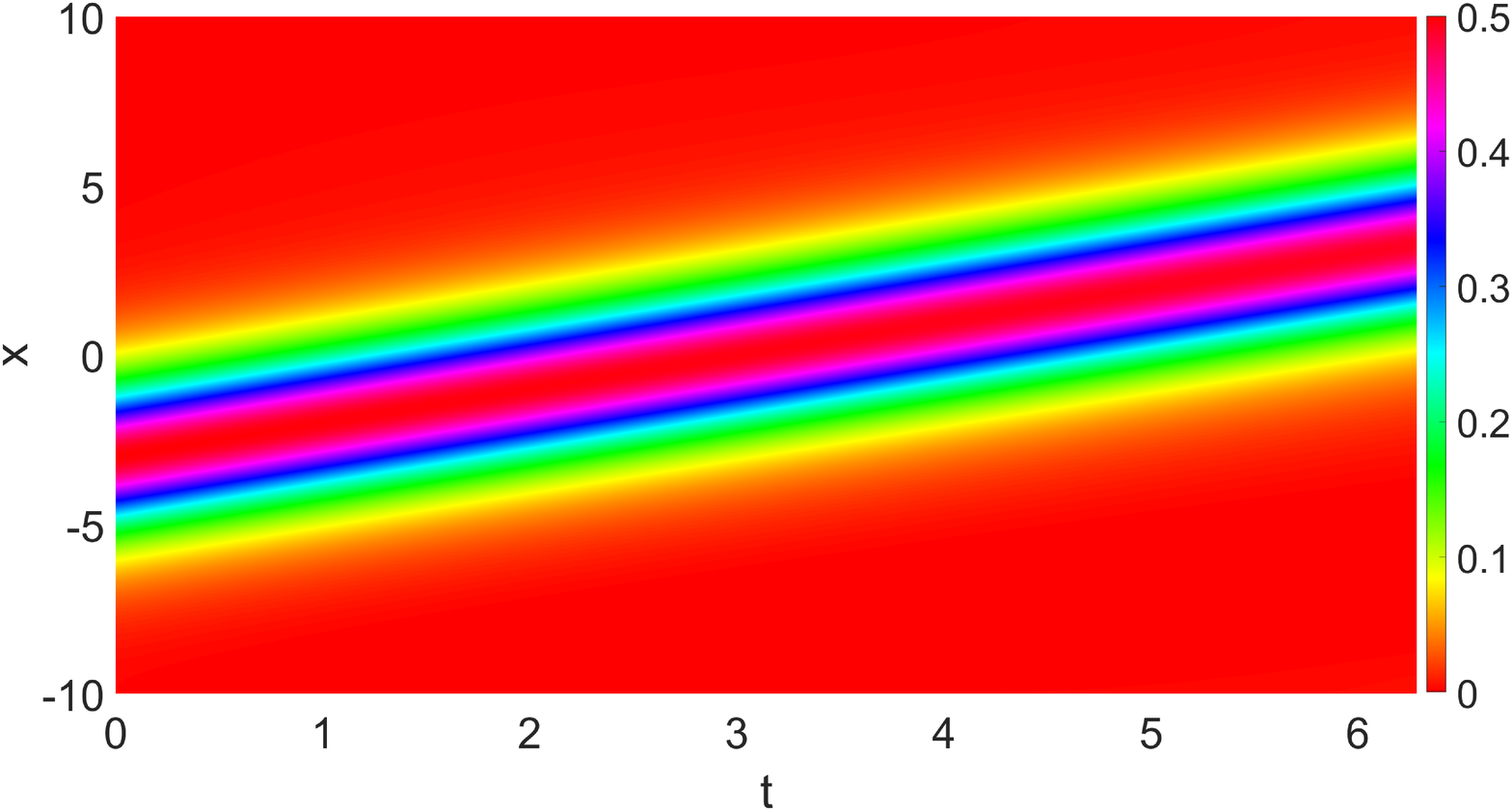}}\\
  \subfigure[$t=1.57$]{\includegraphics[scale=0.3]{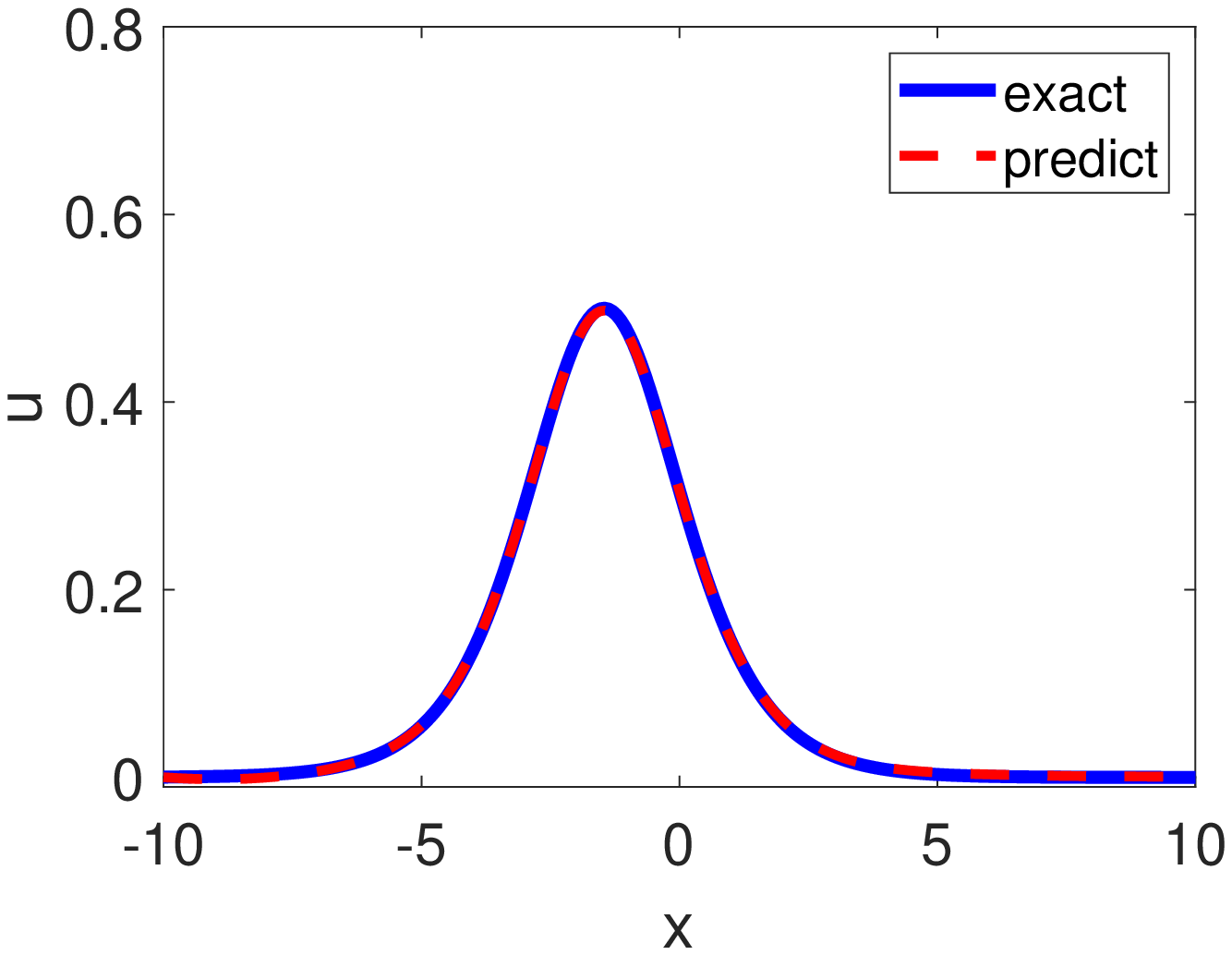}}
  \subfigure[$t=3.14$]{\includegraphics[scale=0.3]{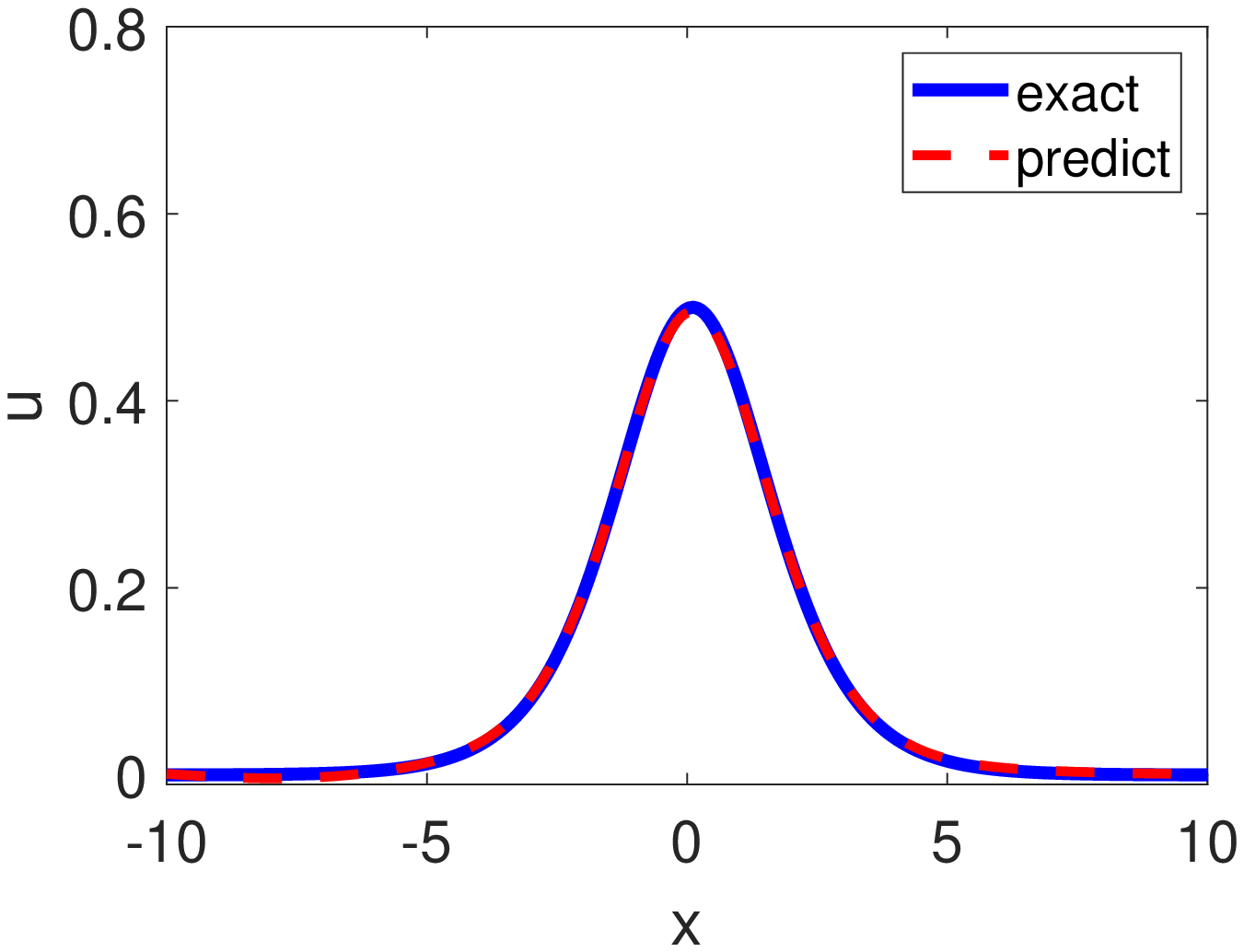}}
  \subfigure[$t=4.71$]{\includegraphics[scale=0.3]{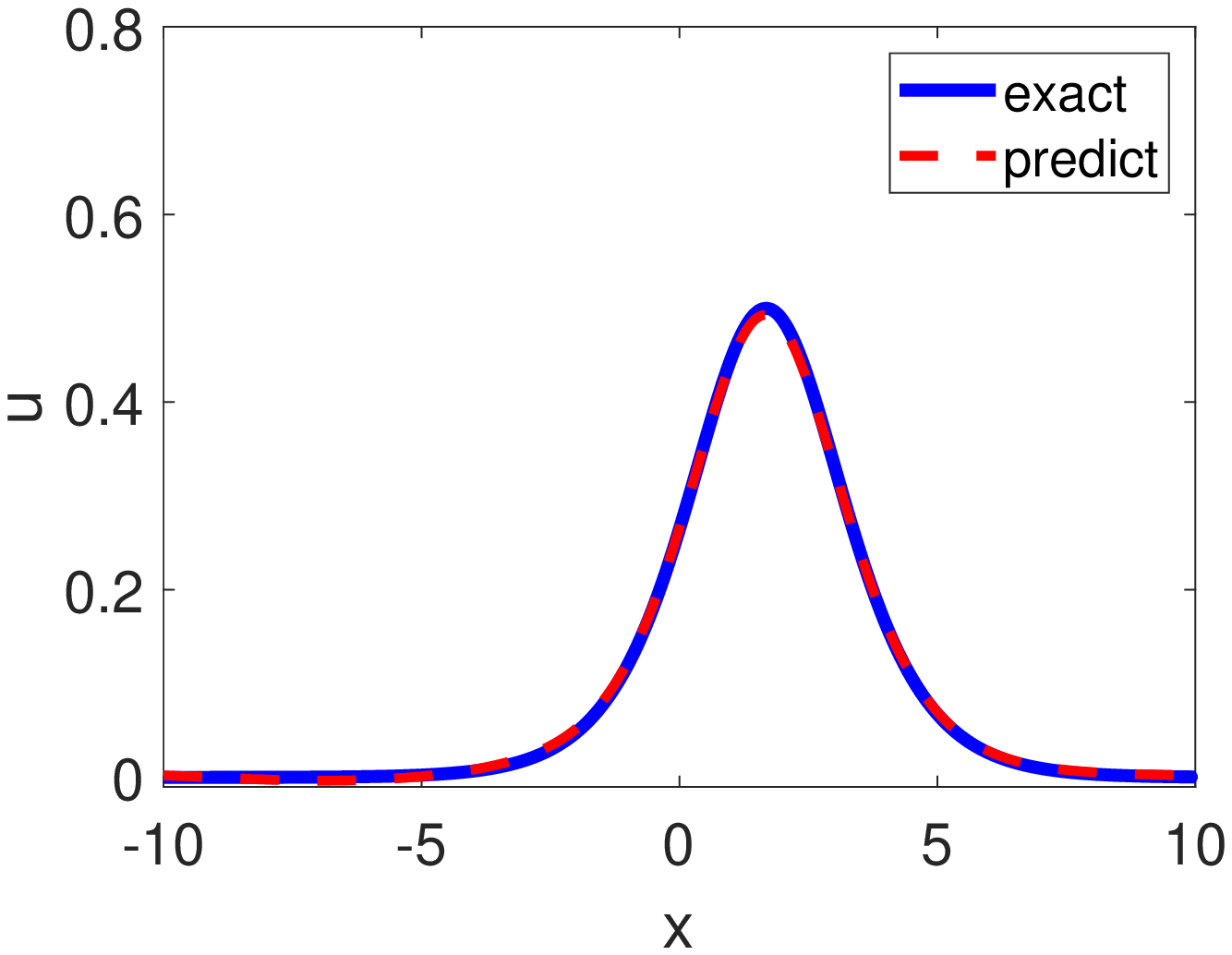}}
  \caption {Subgraph (a) and (b) are comparison of one-soliton exact solution and learned solution of fifth-order KdV equation, subgraph (c)-(e) are the detailed comparison of exact solution and learned spatiotemporal solution at the specific time.}
\label{fig:kdv-1-soliton}
\end{figure}

\begin{figure}[ht]
\centering
\setlength{\abovecaptionskip}{0.cm}
\setlength{\belowcaptionskip}{-0.cm}
  \subfigure[One-soliton Numerical Driven Solution]{\includegraphics[scale=0.35]{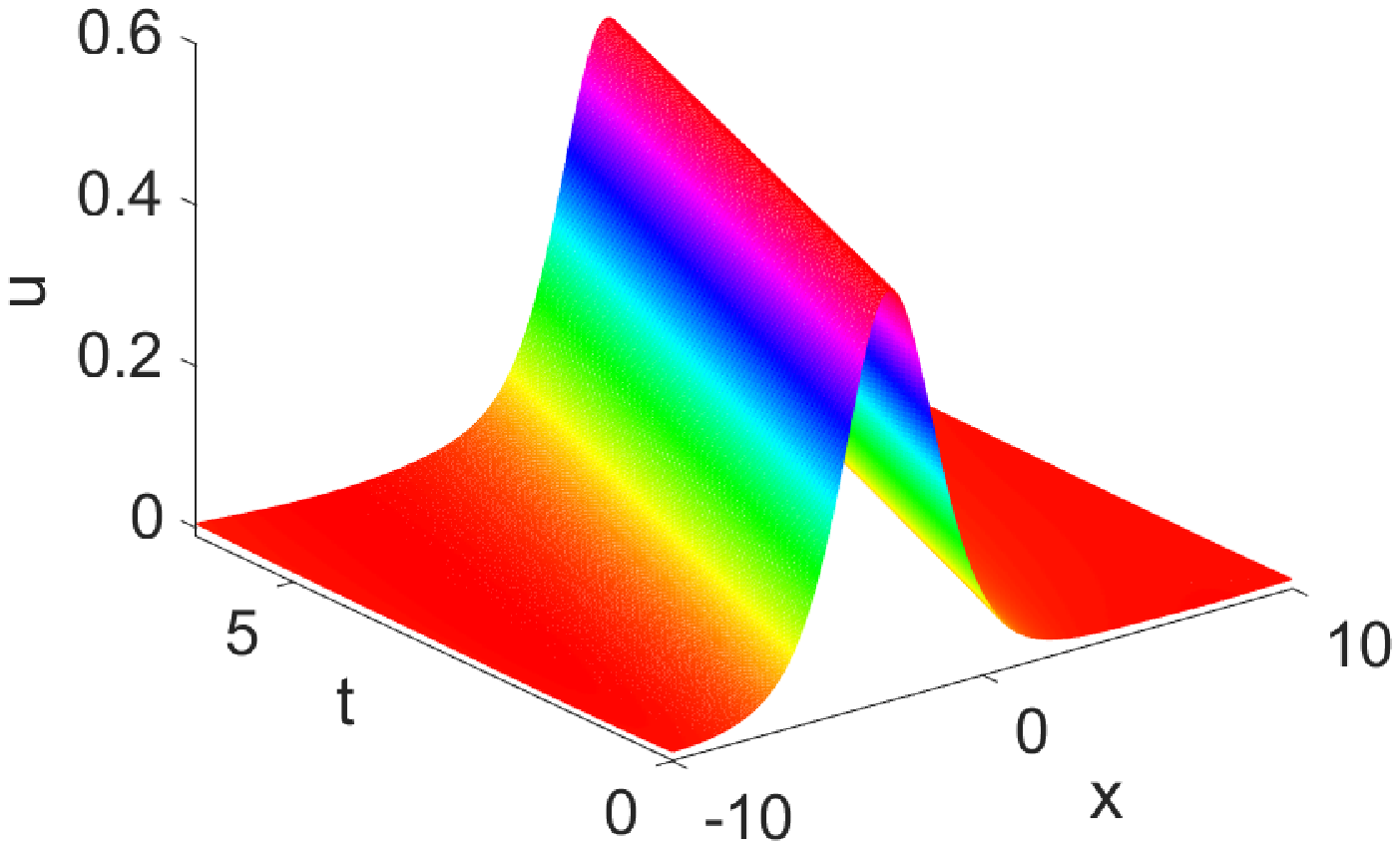}}
  \subfigure[One-soliton Exact Solution]{\includegraphics[scale=0.35]{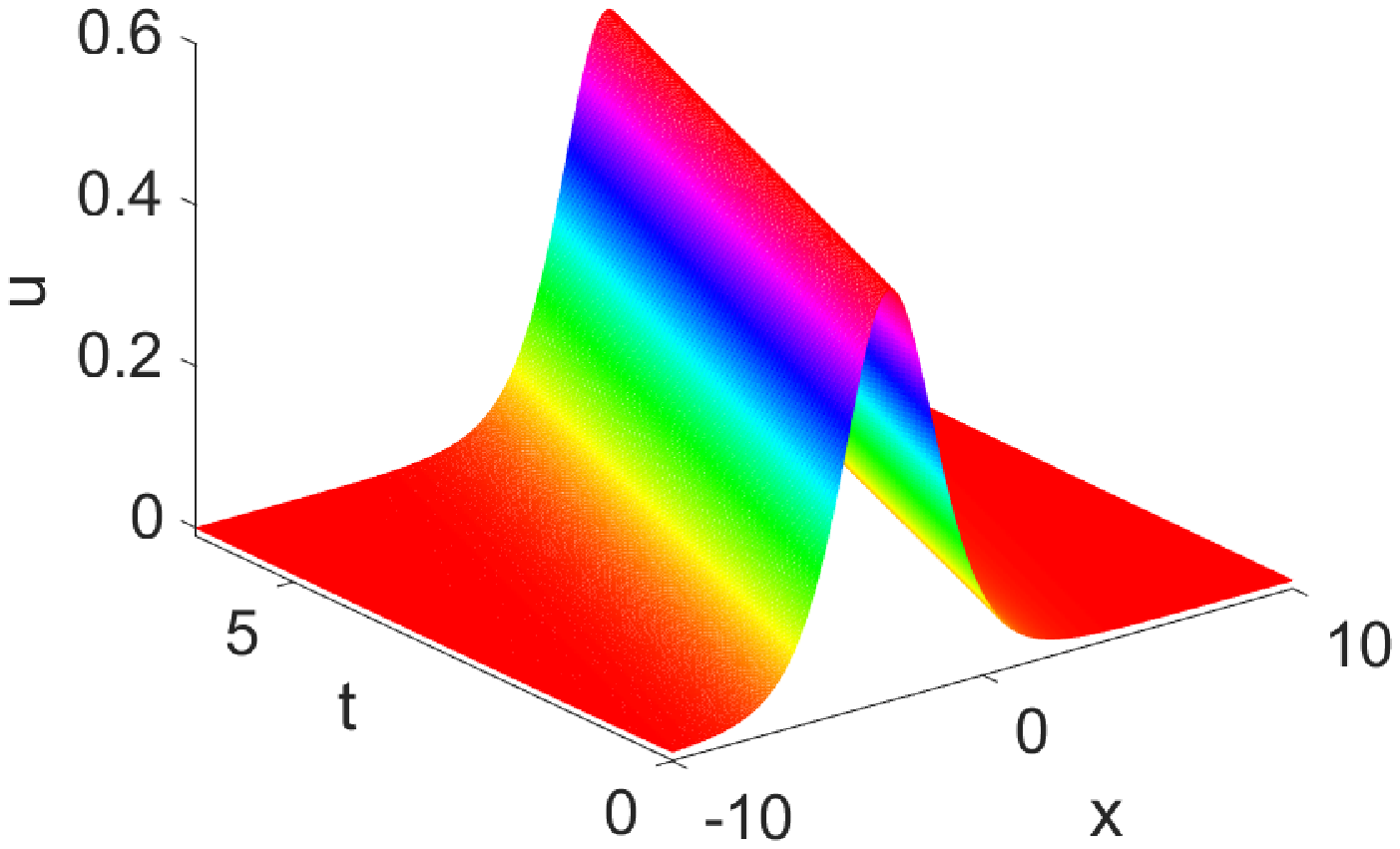}}
  \caption {Spatiotemporal evolution of one-soliton numerical driven solution and exact solution of fifth-order KdV equation}
\label{fig:kdv-1-spatiotemporal evolution}
\end{figure}

In order to obtain high-precision data set, we generate the data of 201 snapshots directly on the regular space-time grid with $\Delta t$ = 0.05s. A small training data subset is generated by randomly latin hypercube sampling method \cite{SteinML}, the number of collection points are $N_u$ = 100, $N_f$ = 20000. Top panel of Figure \ref{fig:kdv-1-soliton} shows the comparison of predicted spatiotemporal solution and exact solution, and bottom panel of Figure \ref{fig:kdv-1-soliton} shows the detailed comparison of exact solution and predicted spatiotemporal solution at different time t=1.57, t=3.14, 4.71 respectively. Figure \ref{fig:kdv-1-spatiotemporal evolution} shows specific spatiotemporal evolution of one-soliton solution of fifth-order KdV equation. The model achieves a relative $L^2$ error of size $1.36 \times 10^{-2}$ in a runtime of 537s. The model is iterated 803 times to complete the operation.

From Figure \ref{fig:kdv-1-soliton} and Figure \ref{fig:kdv-1-spatiotemporal evolution}, we learn that velocity and shape of the numerical driven solution of the one-soliton remain unchanged during the motion, which shows the dynamic behavior of the soliton well.

To verify the universality of our neural network architecture for the one-soliton numerical driven solution of fifth-order KdV equation, we calculate the different one-soliton solution of fifth-order KdV equation.The result shows that the our neural network architecture is very effective in solving one-soliton solution of fifth-order KdV equation.
\begin{table}[H]\footnotesize
\vspace{-0.5cm}
\centering
\caption{The results of different one-soliton solution of fifth-order KdV equation calculated by deep learning method.}
\begin{tabular}{|c| c| c| c |c| c| c| c| }

  \hline
  $k_1$               & 0.9                & 0.95                & 1.0               & 1.05                & 1.1 \\
  \hline
  $L^2$ error         &$1.47\times10^{-2}$ &$2.57\times10^{-2}$  &$1.36\times10^{-2}$ &$4.64\times10^{-2}$ & $2.36\times10^{-2}$\\
  \hline
  time(s)             &687               &680                    &537                 &600                 &1166              \\
  \hline
  Iterations          &300               &803                    &803                 &883                 &1604             \\
  \hline
\end{tabular}\label{tab:kdv-1}
\end{table}
\vspace{-0.4cm}
\subsection{Two-soliton solution}\label{sec:kdv2}
\begin{figure}[ht]
\vspace{-0.4cm}
\centering
\setlength{\abovecaptionskip}{-0.cm}
\setlength{\belowcaptionskip}{-0.cm}
  \subfigure[Two-soliton Interaction of Exact Solution]{\includegraphics[scale=0.14]{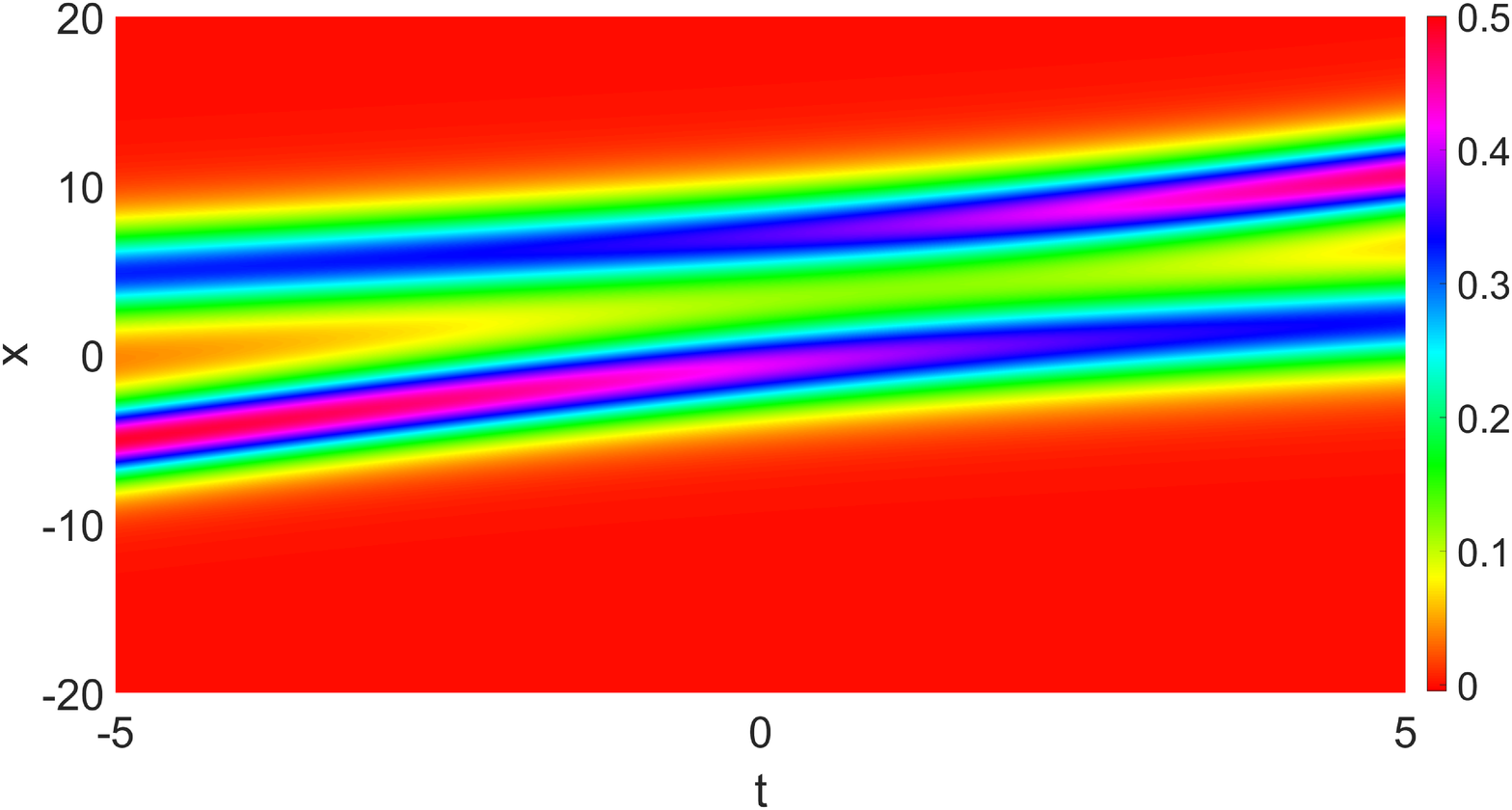}}
  \subfigure[Two-soliton Interaction of Learned Solution]{\includegraphics[scale=0.14]{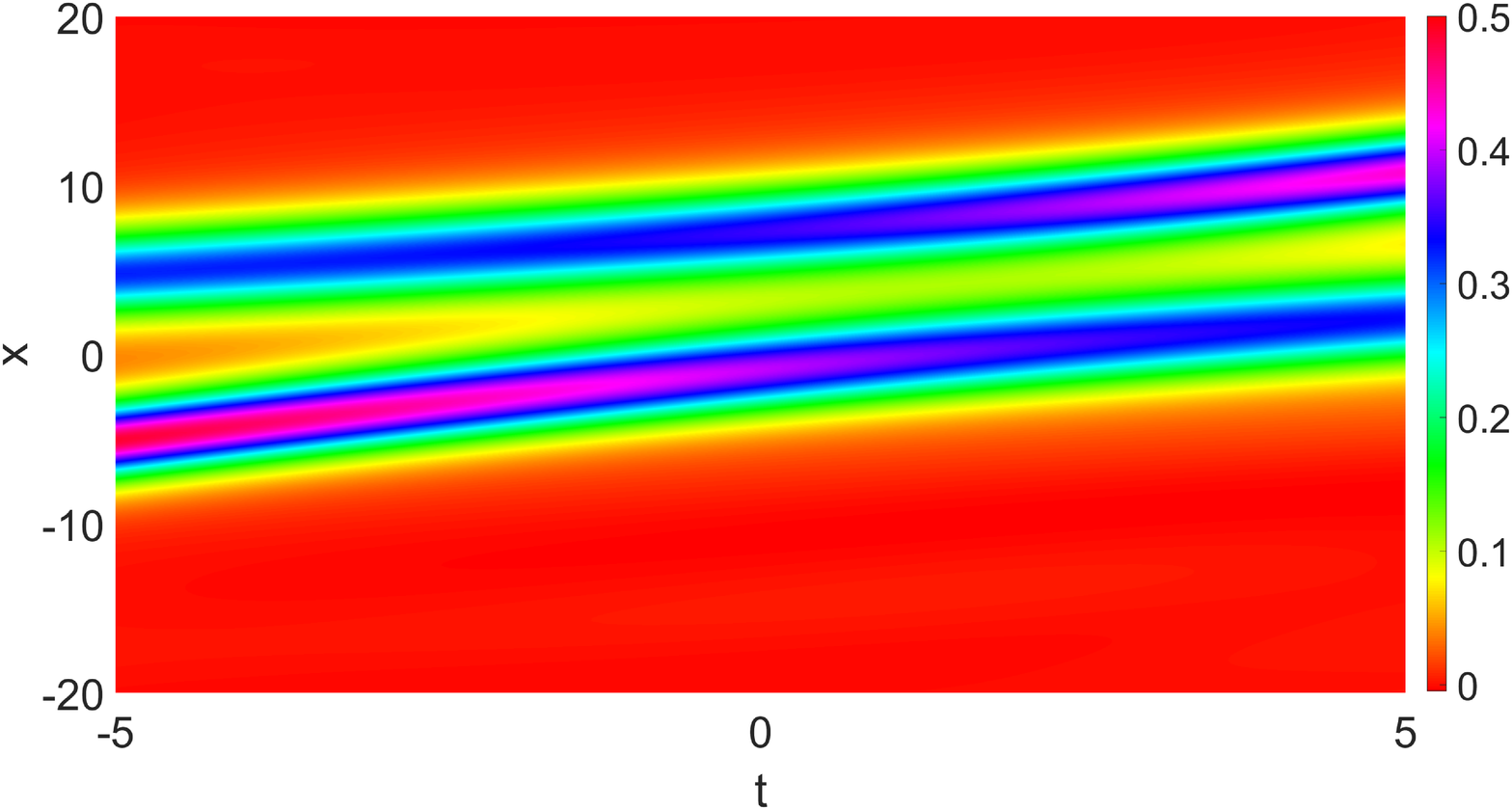}}\\
  \subfigure[$t=-4.5$]{\includegraphics[scale=0.3]{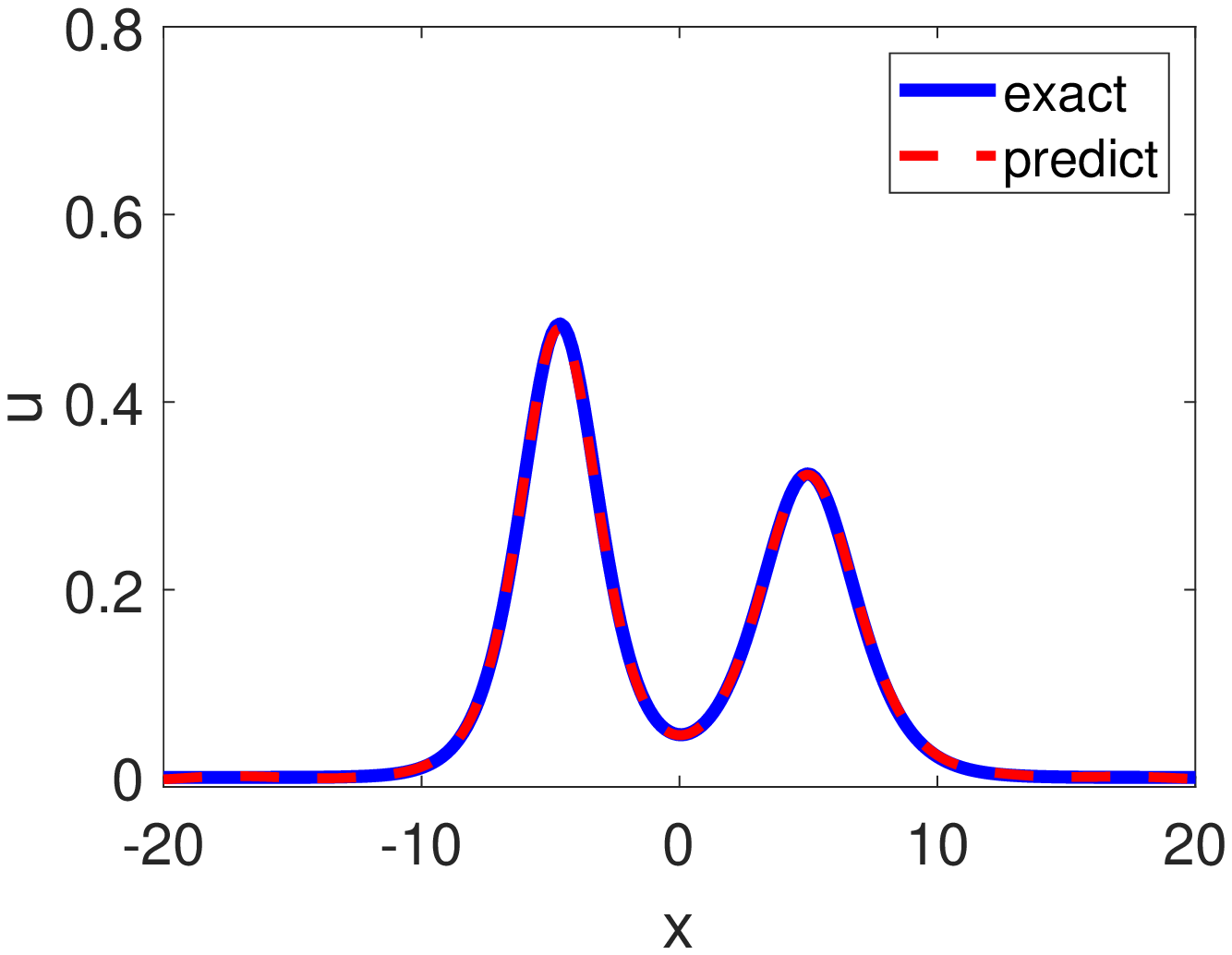}}
  \subfigure[$t=0$]{\includegraphics[scale=0.3]{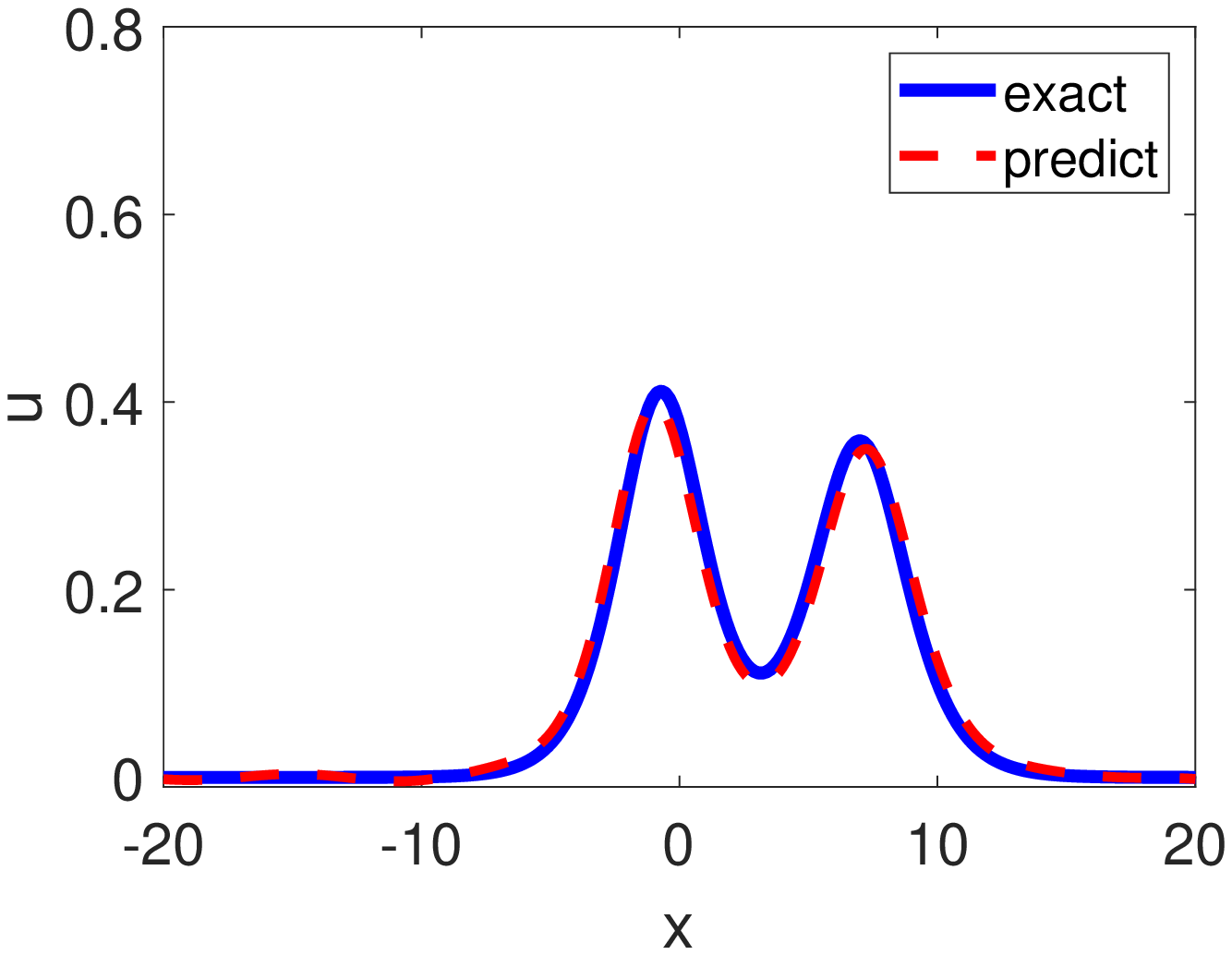}}
  \subfigure[$t=2.5$]{\includegraphics[scale=0.3]{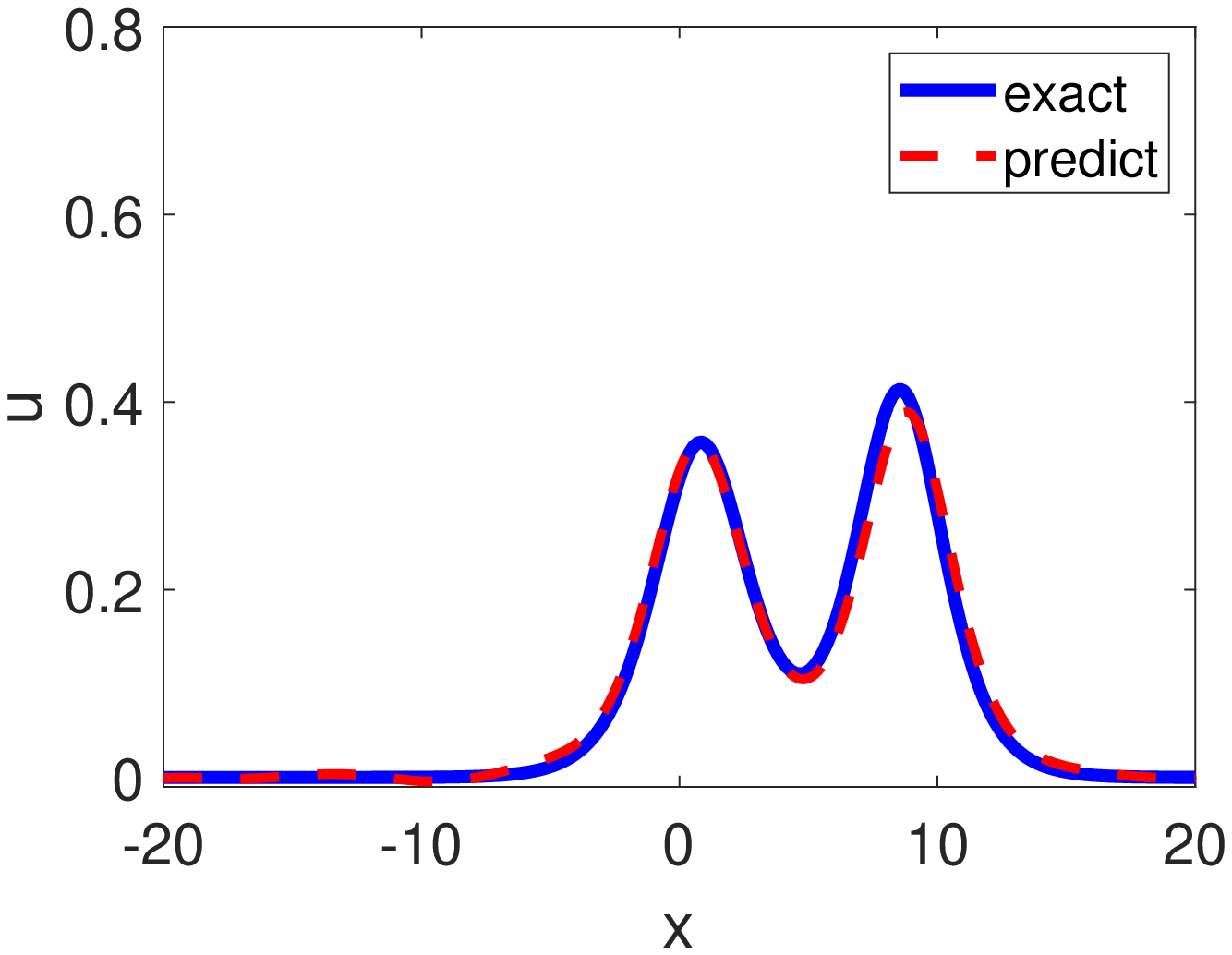}}
  \caption {Subgraph (a) and (b) are comparison of two-soliton exact solution and learned solution of fifth-order KdV equation, subgraph (c)-(e) are the detailed comparison of exact solution and learned spatiotemporal solution at the specific time.}
\label{fig:kdv-2-soliton}
\end{figure}

\begin{figure}[ht]
\vspace{-0.3cm}
\centering
\setlength{\abovecaptionskip}{0.cm}
\setlength{\belowcaptionskip}{-0.cm}
  \subfigure[Two-soliton Numerical Driven Solution]{\includegraphics[scale=0.38]{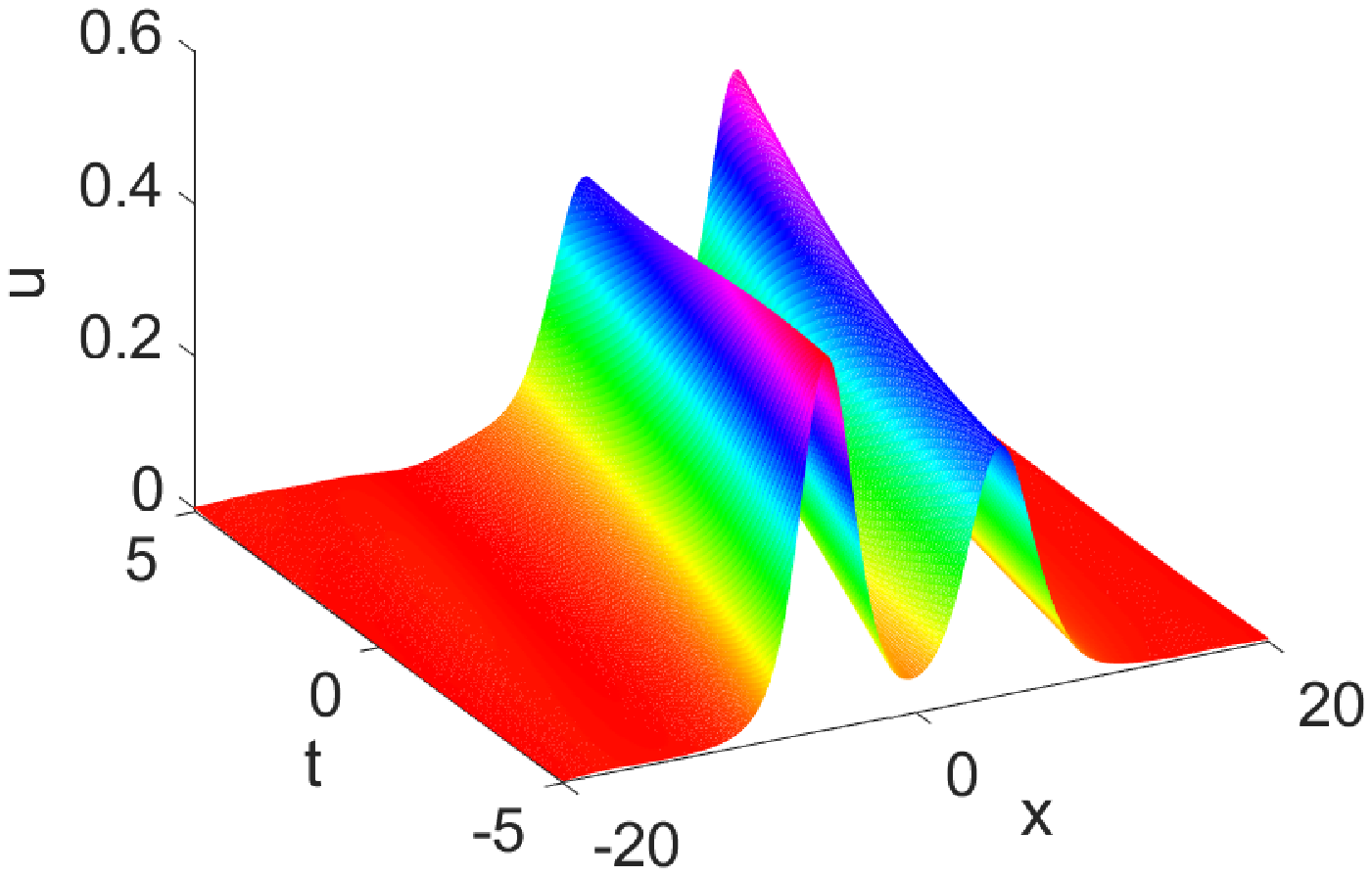}}
  \subfigure[Two-soliton Exact Solution]{\includegraphics[scale=0.38]{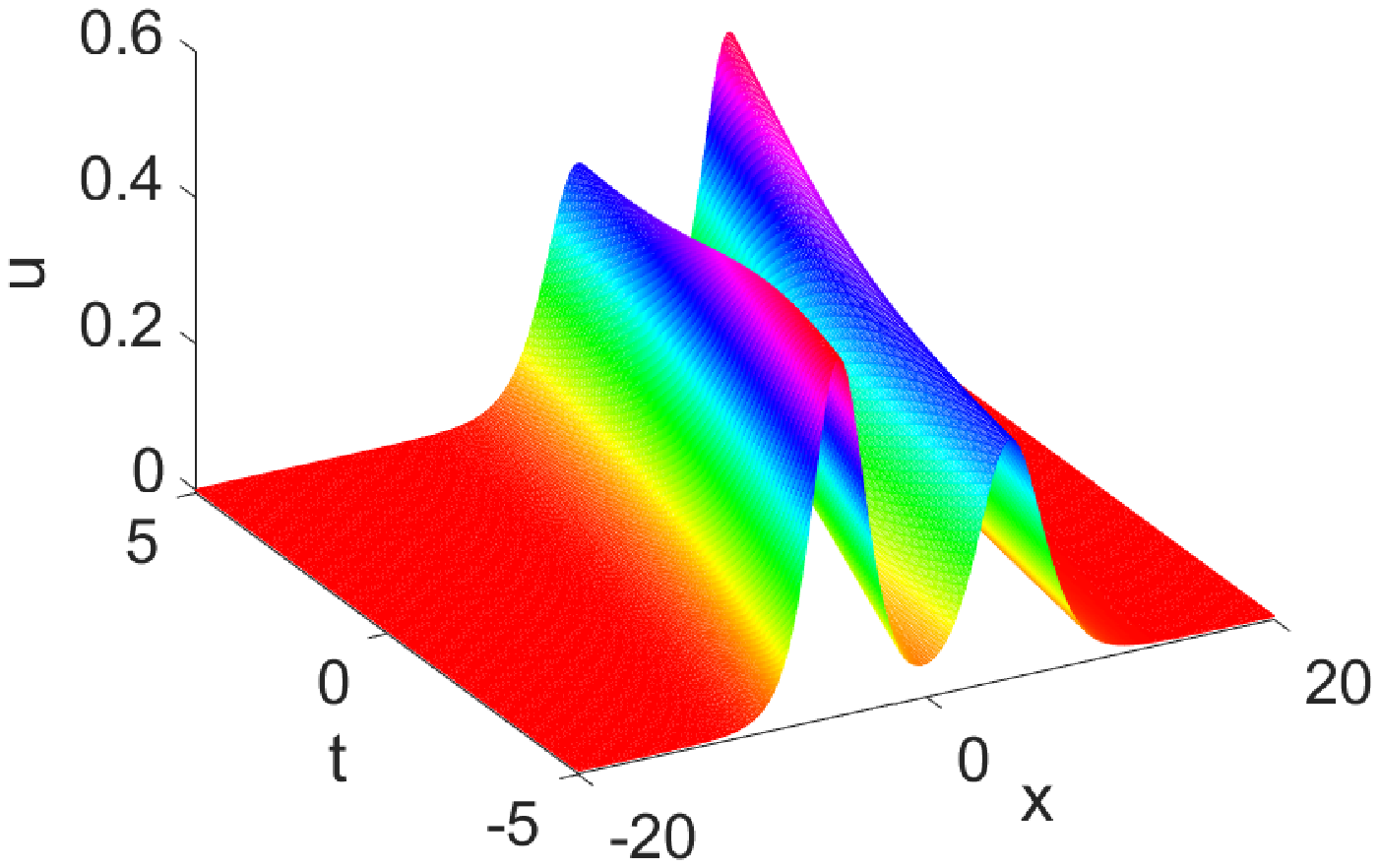}}
  \caption {Spatiotemporal evolution of two-soliton numerical driven solution and exact solution of fifth-order KdV equation}
\label{fig:kdv-2-spatiotemporal evolution}
\end{figure}

In order to observe the solitons interaction behavior well, we set $x\in[-20,20]$ and $t\in[-5,5]$. Equation (\ref{fifth-kdv1}) becomes
\begin{equation}\label{fifth-kdv2}
\left\{
\begin{array}{lr}
u_{t}+(u_{xxxx}+15uu_{xx}+15u^3)_x =0,x\in[-20,20],t\in[-5,5],
\\u(x,0)=u_0(x),
\\u(-20,t)=u(20,t).
\end{array}
\right.
\end{equation}
By using Hirota bilinear method, the two-soliton analytical solution of fifth-order KdV Equation (\ref{fifth-kdv2}) can be obtained\cite{HIETARINTAJ2}
\begin{equation}\label{kdv-02}
\begin{aligned}
u(x,t)=2\frac{k_1^2e^{k_1x+\omega_1t+\delta_1}+k_2^2e^{k_2x+\omega_2t+\delta_2}+(k_1+k_2)^2e^{(k_1+k_2)x+(\omega_1+\omega_2)t+\delta_1+\delta_2+\delta_0}}{1+e^{k_1x+\omega_1t+\delta_1}+e^{k_2x+\omega_2t+\delta_2}+e^{(k_1+k_2)x+(\omega_1+\omega_2)t+\delta_1+\delta_2+\delta_0}} \\-2\frac{(k_1e^{k_1x+\omega_1t+\delta_1}+k_2e^{k_2x+\omega_2t+\delta_2}+(k_1+k_2)e^{(k_1+k_2)x+(\omega_1+\omega_2)t+\delta_1+\delta_2+\delta_0})^2}{(1+e^{k_1x+\omega_1t+\delta_1}+e^{k_2x+\omega_2t+\delta_2}+e^{(k_1+k_2)x+(\omega_1+\omega_2)t+\delta_1+\delta_2+\delta_0})^2},\\
\omega_1=-k_1^5,  \omega_2=-k_2^5,  e^{\delta_0}=\frac{(k_1-k_2)^2+(k_1^2-k_1k_2+k_2^2) }{(k_1+k_2)^2+(k_1^2+k_1k_2+k_2^2)}. \\
\end{aligned}
\end{equation}
We could set $k_1$ = 1, $k_2$ = 0.8, $\xi_1=\xi_2=0$. In order to obtain high-precision data set, we generate the data of 201 snapshots directly on the regular space-time grid with $\Delta t$ = 0.05s. A small training data subset is generated by randomly latin hypercube sampling \cite{SteinML}, the number of collection points are $N_u$= 100, $N_f$= 20000. Top panel of Figure \ref{fig:kdv-2-soliton} shows the comparison of predicted spatiotemporal solution and exact solution. Bottom panel of Figure \ref{fig:kdv-2-soliton} shows the detailed comparison of exact solution and predicted spatiotemporal solution at different time t = -4.5, t = 0, t = 2.5 respectively. Specific spatiotemporal evolution of two-soliton is given in Figure \ref{fig:kdv-2-spatiotemporal evolution}. The model achieves a relative $L^2$ error of size $7.22 \times 10^{-2}$ in a runtime of 1774s. The model is iterated 3606 times to complete the operation.

\vspace{-0.2cm}

On the selection of activation function, we try a lot of other activation functions of two-soliton, the results is given in Figure \ref{tab:compare-kdv}. The result shows that the trigonometric function is more effective in solving the fifth-order KdV equation.
\begin{table}[H]\footnotesize
\vspace{-0.5cm}
\centering
\caption{The results of different activation function of two-soliton calculated by deep learning method.}
\begin{tabular}{|c| c| c| c |c| c| c| }

  \hline
  activation function   & tanh              &cos                & sin                 & sigmoid              & relu                                  \\
  \hline
  $L^2$ error         &$1.46\times10^{-1}$  &$7.22\times10^{-2}$&$1.27\times10^{-1}$ &$8.18\times10^{-1}$  & $6.13\times10^{-1}$ \\
  \hline
  time(s)             &623                 &1740                 & 1531                & 237                  & 48                            \\
  \hline
  Iterations          &1390                 &3600                & 3080                & 0                    & 5                           \\
  \hline
\end{tabular}\label{tab:compare-kdv}
\end{table}
In addition, a small amplitude noise is given to verify the rationality of the structure of the fifth-order KdV equation neural network. The results show that the deep learning method is effective in solving two-soliton solutions of fifth-order KdV equation.
\begin{table}[H]\footnotesize
\vspace{-0.3cm}
\centering
\caption{The results of different two-soliton solution of fifth-order KdV equation calculated by deep learning method.}
\begin{tabular}{|c| c| c| c |c| c| c| c| }

  \hline
  $k_1$               & 1.0                 & 0.99             &1.01              &1.0          & 1.0 \\
  \hline
  $k_2$               & 0.8                 & 0.8              &0.8              &0.79          & 0.81 \\
  \hline
  $L^2$ error         &$7.22\times10^{-2}$ &$8.27\times10^{-2}$ &$1.05\times10^{-1}$&$9.05\times10^{-2}$&$1.0\times10^{-1}$ \\
  \hline
  time(s)             & 1774               &913               &1011              &1000              & 657             \\
  \hline
  Iterations          &  3606              &1654              &1738              &2224              & 1240            \\
  \hline
\end{tabular}\label{tab:kdv-2}
\end{table}
\section{Summary and discussion}\label{sec:summary}
We find a neural network architecture of PINNs suitable for solving high-order nonlinear soliton equation. Specifically, it has four hidden layers, the number of corresponding neurons are 256, 128, 64, 32 and 40, 40, 40, 40. And we study the numerical driven solution of high-order nonlinear problems ( fourth-order Boussinesq equation and fifth-order KdV equation ), and control $L^2$ error to $10^{-2}$ magnitude, which shows that the deep learning method is effective. We extend the deep learning method to the solution of fourth-order and fifth-order equation, but the ability of deep learning method to deal with higher-order equation still needs to be explored.
We summarize the conclusions as follows. Firstly, the deep learning method is suitable to solve the soliton solution of fourth-order Boussinesq equation and fifth-order KdV equation. The deep learning method can recover the dynamic behavior of solitons in high-order nonlinear soliton equation. From numerical driven solution, we can observe the `phase shift' phenomenon, and the shape remains unchanged after the interaction, which is consistent with known facts. Secondly, trigonometric function is effective in high-order nonlinear problems. Compared with the low-order problem, high-order problems have higher sensitivity in selection of architecture of neural network.
\section*{Acknowledgment}
Project supported by LiaoNing Revitalization Talents Program (XLYC1907014) and ``the Fundamental Research Funds for the Central Universities" (DUT21ZD205).

\end{document}